\documentclass[manuscript]{aastex}

\usepackage[letterpaper]{geometry}
\usepackage{apjfonts}
\usepackage{graphicx}
\usepackage{epstopdf}
\shorttitle{Three-dimensional simulations of grain chemistry and ice structure.}
\shortauthors{R. T. Garrod}

\begin{document}

\title{Three-dimensional off-lattice Monte Carlo kinetics simulations of interstellar grain chemistry and ice structure.}

\author{Robin T. Garrod}
\affil{Center for Radiophysics and Space Research, Cornell University, Ithaca, NY 14853-6801,
USA} \email{rgarrod@astro.cornell.edu}

\begin{abstract}

The first off-lattice Monte Carlo kinetics model of interstellar dust-grain surface chemistry is presented. The positions of all surface particles are determined explicitly, according to the local potential minima resulting from the pair-wise interactions of contiguous atoms and molecules, rather than by a pre-defined lattice structure. The model is capable of simulating chemical kinetics on any arbitrary dust-grain morphology, as determined by the user-defined positions of each individual dust-grain atom. A simple method is devised for the determination of the most likely diffusion pathways and their associated energy barriers for surface species. The model is applied to a small, idealized dust grain, adopting various gas densities and using a small chemical network. Hydrogen and oxygen atoms accrete onto the grain, to produce H$_2$O, H$_2$, O$_2$ and H$_2$O$_2$. The off-lattice method allows the ice structure to evolve freely; ice mantle porosity is found to be dependent on the gas density, which controls the accretion rate. A gas density of $2\times 10^{4}$ cm$^{-3}$, appropriate to dark interstellar clouds, is found to produce a fairly smooth and non-porous ice mantle. At all densities, H$_2$ molecules formed on the grains collect within the crevices that divide nodules of ice, and within micropores (whose extreme inward curvature produces strong local potential minima). The larger pores produced in the high-density models are not typically filled with H$_2$. Direct deposition of water molecules onto the grain indicates that amorphous ices formed in this way may be significantly more porous than interstellar ices that are formed by surface chemistry.

\end{abstract}


\section{Introduction}

Chemistry on dust-grain surfaces is crucial to the chemical evolution and enrichment of the interstellar medium. The simplest and most abundant interstellar molecule, H$_2$, has long been known to form exclusively on dust-grain surfaces (Gould \& Salpeter 1963) before rapidly returning to the gas phase, where it participates in ion-molecule reactions that may increase the chemical complexity of the gas phase (Dalgarno \& Black 1976). However, in both quiescent clouds and in protostellar envelopes, the accretion onto the grains of gas-phase atoms and molecules also leads to the build-up of molecular ice mantles (e.g. Whittet et al. 1989; Chiar et al. 1995; Gibb et al. 2000; Boogert et al. 2008). The observed dust-grain ices are composed primarily of water ice (H$_2$O), as well as other simple hydrides such as methane (CH$_4$) and ammonia (NH$_3$), and other organic species including carbon monoxide (CO) and carbon dioxide (CO$_2$). These ices are understood to form primarily at low dust temperatures ($\sim$10 -- 20 K), with the hydrides forming mainly by the surface diffusion and repetitive addition of atomic hydrogen to other atoms (e.g. Hasegawa et al. 1992). Atomic H remains significantly mobile even at low temperatures (e.g. Manic{\`o} et al. 2001), which allows it to diffuse around the surface of the grain (or the surface of the ice mantle), thermally hopping from one binding site (i.e. potential well) to another, until it meets a reactive species and forms a new molecule. Solid-phase water, thus formed, may account for the majority of the oxygen budget in many dense interstellar regions.

The porosity of both the dust-grain ices and of the grains themselves has recently been suggested as an important influence on grain-surface chemistry. Chemical models that treat the formation of H$_2$ within pores in the grains (Perets \& Biham 2006) and the formation of other molecules in pores within the ices (Taquet et al. 2012) have demonstrated significant -- albeit moderate -- effects in each case. These models combine rate equation-based treatments of the chemical kinetics with generic assumptions regarding the characteristics of the pores. Experiments on ice mixtures (Collings et al. 2004; Fayolle et al. 2011) have also shown that pores in laboratory water ice mixtures may absorb more volatile species, such as CO and CO$_2$, to be released at higher temperatures. Amorphous solid water (ASW) ices formed in the laboratory by deposition of water molecules directly onto a cold substrate show a large degree of microporosity, defined as pores with diameter $<$20 \AA\, (Raut et al. 2007 and references therein); however, it is unclear whether ices formed by such means are representative of the structure of interstellar dust-grain ices formed largely by surface-chemical processes over many thousands of years.

The large-scale morphology of interstellar dust grains is also likely to be important to the ice chemistry and structure. Interplanetary dust grains appear to be irregular in shape, and interstellar dust models imply ``fluffy'' morphologies (Mathis 1996); interstellar dust coagulation simulations also appear to confirm this view (Ossenkopf 1993; Ormel et al. 2009). The chemical implications of such morphologies cannot be investigated using current dust-grain chemical models, which do not specify actual grain shapes. 

Various models currently exist for the simulation of interstellar dust-grain surface chemistry, which may be divided into broad two groups. The most common is that which uses generalized reaction rates for surface species, based on representative values for the binding energies and surface-diffusion barriers for each atom or molecule. This chemical system may be simulated using either numerical solver routines to integrate ordinary differential equations (e.g. Hasegawa et al. 1992; Garrod \& Herbst 2006; Garrod 2013), or using Monte Carlo (MC) techniques (Charnley 2001; Vasyunin et al. 2009; Vasyunin \& Herbst 2013). However, in either case, little structural information about the ice can be obtained, because the {\em positions} of atoms and molecules are not explicitly modeled, other than (in some cases) by depth. Furthermore, the effects of local binding properties of the surface (caused by its shape or molecular composition) cannot be accounted for by this approach; the models implicitly assume a perfectly smooth grain with uniform binding. Unfortunately, heterogeneous surface effects appear to be important, even in very simple systems such as that of H$_2$ formation on the simplest non-smooth surface (Cuppen \& Garrod 2011).

The other type of model uses a microscopically-exact kinetic Monte Carlo approach (Chang, Cuppen \& Herbst 2005; Cuppen \& Herbst 2007; Cuppen et al. 2009), which allows the behavior of each individual surface particle to be simulated. These models begin with a square section of grain surface, with periodic boundary conditions to imitate the closed surface of a dust grain. Using a so-called {\em on-lattice} approach, atoms and molecules on the surface are allowed to diffuse between pre-defined lattice positions, with diffusion barriers and binding energies dependent on the pair-wise interations between species in adjacent binding sites. Every diffusion, desorption, accretion and reaction event is treated explicitly; consequently, the method is more computationally demanding than those based on simple rate equations, and requires a simpler network of species and reactions.

While the latter method produces valuable structural information, it retains several limitations. The use of an on-lattice approach means that the surface species are not fully free to adapt their positions to local conditions, and the sizes of individual molecules are not taken into account in any way. Furthermore, the method does not treat chemistry on a three-dimensional grain surface, and so the effects of large-scale grain morphologies on the surface chemistry and ice structure cannot be investigated.

An {\em off-lattice} approach requires information concerning the possible diffusion pathways of a surface particle, as well as its final position, as these considerations are no longer guided by the fixed lattice. Off-lattice techniques have been used in other fields (e.g. Konwar et al. 2011, and references therein), using pre-calculated, generic diffusion pathways. Such information may be obtained through intensive molecular dynamics calculations, and incorporated into kinetic Monte Carlo models of diffusion on regular surface structures. However, for an amorphous ice surface, composed of multiple different chemical species, the number of calculations and data to be stored using such a method would be extremely large. The long timescales involved in interstellar chemistry would also likely make such intensive approaches prohibitively slow.

Here, a new off-lattice astrochemical kinetic Monte Carlo model of interstellar grain-surface chemistry is presented. A simple method is devised for the {\em ad hoc} calculation of the post-hop positions of diffusing particles, under the assumption that thermal hopping occurs through easily-identified surface-potential saddle points. Particle positions are determined by local minima in the surface potentials felt by individual particles, as defined by the binding partners of each surface species. In this way, the positions and chemical kinetics of every particle on a dust grain may be traced as atoms accrete from the gas phase and an ice mantle is formed. The model allows chemistry to be simulated on a three-dimensional dust grain of arbitrary size and shape, whose surface is explicitly defined by the positions of its constituent atoms. The morphology and porosity of the resultant ice structure is determined purely by the chemical kinetic processes occurring on the grain/ice surface.

Results are presented for a preliminary version of the new model, as applied to a range of gas densities, using a simple chemical network and a small, idealized dust grain. Future work will consider the parameter space in greater depth.

Section 2 details the new method; the results are presented in Section 3. A discussion of the model and its initial results are provided in Section 4. The conclusions of this work are outlined in Section 5.

\section{Method: Monte Carlo Simulations}

The model presented here is used to trace the microscopic chemical kinetics that occur on an interstellar dust-grain/ice surface as the result of the accretion and diffusion of atoms derived from the gas phase. The model starts with a three-dimensional dust grain (see Fig. 1), whose surface is explicitly defined by the positions of the constituent atoms. The arrival of a particle from the gas phase results in its binding in a local potential minimum, whose position and strength are determined by the combined pair-wise interaction potentials between the particle and the nearest surface atoms. The number of surface binding partners is dependent on the morphology of the surface, including the presence of other accreted surface particles. The strength of the pair-wise interaction potential between specific chemical species is defined explicitly within the model (see Table 1). Binding is treated as a  "physisorption" or van der Waals-type interaction. No {\em chemical} bonding to the grain surface itself is considered.

In general, a surface particle may undergo thermal desorption into the gas phase, or may diffuse to a binding site (i.e. surface potential minimum) adjacent to its current location, via thermal hopping. The interaction potentials felt by the particle determine the rates at which such processes may occur, and thus determine the competition with similar processes for other surface species or with further accretion from the gas phase. Details of each of these model elements are presented in the subsections below.

In this preliminary model, all atoms and molecules -- including atoms that constitute the dust grain itself -- are assigned a uniform radius, with an assumption of spherical symmetry. Experimentally-determined charateristic radii for water molecules, which comprise the majority of the dust-grain surface ice in the models presented here, are dependent on the overall ice structure. Measured values for the intermolecular distance in amorphous solid water (ASW) ice can range to as high as 3.3 \AA \,(Angell 2004), while static, non-polarizable water models produce values around 3.16 \AA \,(Vega et al. 2009), with radii centered on the oxygen atom. An intermolecular spacing of $\sigma=3.2$ \AA \,is chosen for the current model, which for practical purposes defines the hard-sphere radius of a water molecule to be 1.6 \AA; this value is applied to all other chemical species. Canonical van der Waals radii for oxygen and hydrogen atoms are 1.52 \AA \, and 1.20 \AA , respectively (Bondi 1964). The choice of a water-specific generic value ensures that the sizes of the dust-grain ice mantles produced by the models are appropriate. Future models will incorporate van der Waals radii from the literature, allowing the ice structure to reflect more accurately the varying sizes of the constituent atoms and molecules.

\subsection{Numerical method}

The model adopts the numerical Monte Carlo (MC) approach suggested and proved by Gillespie (1976), and outlined below; the general method has been successfully used in the past for astrochemical simulations by several authors (e.g. Charnley 2001; Chang, Cuppen \& Herbst 2005, Cuppen \& Herbst 2007, Vasyunin et al. 2009). In common with the work of Cuppen, the new model takes a microscopic view of the grain-surface chemistry, but otherwise uses a new methodology for the treatment of the chemistry and the physical positions of atoms and molecules on the grain.

Gillespie (1976) treats the time evolution of a chemical kinetic system as a series of consecutive events. The two key quantities to be determined by the probabilistic approach at any time, $t$, are: (i) which event occurs next, and (ii) how much time elapses between this and the previous event. To determine this information, each possible chemical process, $i$, is assigned a rate, $R_{i}$ (in units of s$^{-1}$). In the new model, such processes include an accretion event, a desorption event, or an individual thermal hop from one binding site to another; in the case of desorption and diffusion, specific rates are calculated for each individual surface particle. Following Gillespie, each process is then assigned a probability equal to its own rate divided by the sum of all rates, $R_{\mathrm{tot}}$. A computer-generated random number,  $0 \leq N_{\mathrm{ran,1}} < 1$, then chooses from these probabilties which process will occur next. Another random number,  $N_{\mathrm{ran,2}}$, determines the time elapsed since the last event, equal to $\Delta t = -\ln (N_{\mathrm{ran,2}}) / R_{\mathrm{tot}}$. The populations, positions, rates, or any other relevant information for each affected chemical species is then updated and the above steps are repeated until the user-defined end-time of the simulation is reached. In the model, this method is used to determine the overall sequence of thermal hopping, reaction, desorption or accretion of every surface atom/molecule, according to their individual rates.

\subsection{Generation of an idealized interstellar grain}

The new model takes explicit account of the positions of particles on the grain surface, without using a pre-determined position lattice. However, it {\em does} require the grain-surface structure to be defined initially. The model will allow chemical simulations on any arbitrary grain size and structure.

For this purpose, a simple code has been constructed to produce coordinates for every atom in an approximately spherical dust-grain of radius 5 atoms; see Fig. 1. The grain may be simplistically considered to be composed of carbon atoms, although the purpose of the generated grain is to provide a simple surface structure upon which the new model may be initially tested. Atoms are positioned in a simple cubic structure, and assigned the representative radius of 1.6 \AA\, (which varies only slightly from the canonical van der Waals radius for carbon atoms of 1.70 \AA; Bondi 1964). The grain has a diameter of approximately 32 \AA. Future applications of the generating code will be used to produce alternative/arbitrary grain sizes and morphologies, as well as more specific surface structures, compositions, and atomic sizes.

\subsection{Accretion of gas-phase species}

The total rate of accretion of gas-phase material onto the grain is dependent on the cross-section presented by the grain and its surface ice mantle. In simpler models, this rate is typically approximated by assuming a spherical grain, using a mean radius corresponding to either a bare grain (Hasegawa et al. 1992) or the combined grain and mantle (Acharyya et al. 2011). 

However, unlike most past models, the new method explicitly tracks the position of each surface species, and allows the non-uniform build-up of surface structures. Thus, the treatment of accretion adopted in this model must account for the {\em trajectory} of the incoming particle as well as the rate.

In the new model, accretion from the gas phase is handled in several stages; firstly, a basic accretion rate is constructed for each gas-phase chemical species, based on the cross-section of the smallest sphere that can contain the entire grain and ice mantle. Following Hasegawa et al. (1992), 
\begin{equation}
R_{\mathrm{acc}}(i) = \sigma_{\mathrm{sphere}} \, <v(i)> \, n(i)
\end{equation}
for the sphere bounding a single dust grain, where $v(i)$ and $n(i)$ are the thermal velocity and concentration of the accreting gas-phase species, $i$, and $\sigma_{sphere} = \pi \, r^{2}$, where $r$ is the radius of the sphere. (Because the bounding sphere is by definition larger than the combined mantle and dust grain, this basic rate is larger than the effective rate ultimately produced by the model, as explained below). An independent accretion rate for each gas-phase species is constructed in this way.

At such time that the Monte Carlo routine selects, according to the rate given by equation (1), the accretion of a particular species to be the next process in the sequence, a randomized entry point into the bounding sphere is then generated in spherical coordinates:
\begin{eqnarray}
\theta = 2 \, \pi \, N_{\mathrm{ran,3}}\\
\phi = \cos^{-1} \, (2 \, N_{\mathrm{ran,4}} -1)
\end{eqnarray}
where $N_{\mathrm{ran,3}}$ and $N_{\mathrm{ran,4}}$ are random numbers between 0 and 1. Equations (2) and (3) result in a uniform distribution of points over the surface of the bounding sphere.

Expressions of the same form as equations (2) and (3), using new random numbers, are then used to determine a trajectory for the incoming particle that is randomized over solid angle. However, only {\em inward} trajectories are considered (outward trajectories are reversed to point inward), thus avoiding the double-counting of cases where a particle enters from an alternative point on the sphere. 

The randomly-generated trajectory is traced through the interior of the sphere until a grain/ice particle is met by the incoming particle. At this stage, the accretion is judged to be successful, and another routine finds the nearest potential well to the contact point, assuming Lennard-Jones potentials for all interactions between the accreting particle and the grain/ice particles (see section 2.3.2).

If the trajectory of the incoming particle does not intersect with any atoms/molecules on the grain or surrounding ice mantle, then the accretion was unsuccessful. In such a case, the clock is nevertheless advanced, and the Monte Carlo selection process continues as usual. The entry of a particle into the bounding sphere is still a valid process, even if no accretion ultimately results. The choice of the radius of the bounding sphere could be made arbitrarily large (producing very low accretion efficiency for particles entering the bounding sphere), with no loss of accuracy to the model results. The purpose of the adoption of the smallest bounding sphere possible is simply to avoid unnecessary and time-consuming calculations.

In the results presented, only atomic H and O are allowed to accrete from the gas phase, with an assumed sticking coefficient of unity. Fractional abudances of $2 \times 10^{-4}$ are adopted for each species. The overall gas density is varied in the models, but the fractional abundances are assumed to remain constant over time, and no gas-phase chemistry is treated in the current model. Each of these restrictions may be lifted in future work.

\subsection{Grain-surface processes}

Monte Carlo treatments of {\em gas-phase} chemistry need consider only the evolution of, and rates associated with, the {\em total} population of each gas-phase chemical species. However, a microscopically-exact treatment of ice structure requires the positions and behavior of individual grain-surface atoms and molecules to be simulated explicitly.

In this model, rates for diffusion and desorption are calculated for each individual grain-surface atom or molecule. Diffusion and desorption are treated as thermal processes, with individual rates of the form:
\begin{equation}
R(j) = \nu_{j} \, \exp (-E_{j}/T)
\end{equation}
where $j$ is a specific atom or molecule, $\nu_{j}$ is the characteristic vibrational frequency of the species in question, $T$ is the dust temperature, and $E_{j}$ is either the binding energy or diffusion barrier, depending on which process is being considered. In this paper, the terms ``thermal desorption'' and ``evaporation'' are considered synonymous.

In practice, the Monte Carlo procedure initially selects a {\em particle} to undergo a thermal process, using the rate of the fastest thermally-activated process allowed for each surface particle in order to make the selection (although these rates also compete with the accretion rates described above). From here, a random selection is made between desorption and each and every diffusion process that the chosen particle may undergo, weighted according to each individual rate. This two-stage approach is necessary because, for example, the probability that a particle attains sufficient energy to diffuse also contains the small probability that it attains sufficient energy to desorb entirely; equation (4) represents the rate at which a particle reaches an energy of $E_j$ {\em or higher}. These processes are therefore in competition with one-another and must be treated as such. In the same way, for the case of a particle that has, for example, four available diffusion paths with equal energy barriers, the diffusion rate is not increased four-fold, but split four ways. 

The atoms that constitute the dust-grain surface itself are not allowed to diffuse or desorb, but do contribute to the binding energies of surface-bound species.

The treatments applied to desorption and diffusion processes are described in more detail below.

\subsubsection{Thermal desorption}

When an atom has just accreted onto the grain surface, it initially resides in a potential well whose position and binding energy are calculated at the time of accretion. (It should be noted that the position assigned to each particle represents only its mean position; in reality, all particles would be vibrating in their potential wells). In the model, a particle is considered to be bound to a surface if it is bonded to 3 or more other atoms/molecules -- these could be atoms of the dust-grain itself, or other particles that are themselves bound to the grain surface. Two particles, A and B, are considered to be bonded to each other if their center-to-center separation, $s_{\mathrm{AB}}$, is approximately equal to $\sigma = 3.2$ \AA \, (specifically, in the range $0.9 \sigma < s_{\mathrm{AB}} < 1.1 \sigma$; closer bonding is explicitly prohibited in the model, in consideration of van der Waals repulsion).

For each individual atom/molecule, the binding energy (i.e. the energy required to desorb/evaporate into the gas phase), $E_{\mathrm{des}}$, is equal to the sum of the interactions, $\epsilon$, with the other particles to which it is bound (N.B. equation (5), below, is {\em not} used for these calculations). This value is used to evaluate the evaporation rate given by Eq. (4), and in the calculation of $\nu_{j}$, following Hasegawa et al. (1992). Any particles that are spatially  ``boxed-in'', under the hard-sphere approximation, are not allowed to evaporate or diffuse. Thus, molecules within bulk ice mantles are fixed in place unless the outer-surface molecules are removed.

Interaction potentials are defined for the pairing of every type of chemical species in the model, as shown in Table 1. Potentials between any chemical species and a particle in the dust grain itself are chosen to reproduce approximately the binding energies used in previous, rate-based gas-grain chemical models, under the assumption that binding to the bare grain involves bonding with 4 dust-grain atoms. For example, the binding energy of atomic hydrogen assumed by Garrod \& Herbst (2006) of 450 K may be compared with a value in this model of 400 K, which would be obtained for bonding to a flat face of the dust grain used in the model (see Figs. 1 \& 2). Bonding to only three grain atoms, or to as many as 5, is possible in some positions on the bare grain surface, which would produce a binding energy of 300 K, or 500 K, respectively. For atomic oxygen, the 200 K interaction potential with a dust-grain atom adopted in this model reproduces the commonly-used binding energy of 800 K (Tielens \& Allamandola, 1987), for most positions on the grain.

Interaction potentials between particles that are not a part of the grain itself are nevertheless guided by the grain-binding values. In general, the interaction potential of a pairing A--B is assigned the lower of the values for A--gr and B--gr (where ``gr'' represents a dust-grain atom). An exception is permitted for species expected to demonstrate strong hydrogen-bonding properties. Thus, the interaction potentials involving H$_2$O, H$_2$O$_2$, and OH (when bound to one of those two molecules) are augmented by several hundred Kelvin. For H$_2$O--H$_2$O and H$_2$O$_2$--H$_2$O$_2$ bonding, the A--gr values are doubled. This produces a 4-partner binding energy for water of 4000 K, although in practice the model may yield water-ice that is sufficiently amorphous to produce many surface water molecules that are bound to 5 or 6 other waters. The generic value adopted by Garrod \& Herbst (2006) was 5700 K.

The binding treatment assumed throughout the model may be characterized as a ``hard-sphere, nearest-neighbor'' approach. The interaction potentials assumed in the model are isotropic; polar molecules may in reality exhibit more directionality, but such considerations are currently outside the scope of this model.

\subsubsection{Diffusion}

Because the model uses an ``off-lattice'' approach, the positions of the adjacent potential wells into which a particle could diffuse are not initially specified, and must be calculated -- as needed -- from the current physical state of the system, as defined by the positions of all nearby particles. The method set out for this model allows the energy barrier for each possible diffusion pathway of a given surface particle to be calculated at the same time as the total binding energy, prior to any calculation of a final {\em post-diffusion} position associated with it. The putative final position resulting from a thermal hop need therefore only be calculated if that particular hop has actually been selected as the next process.

To determine diffusion barriers, the model requires information about the possible {\em pathways} that a diffusing particle may take from the current binding site to another. For the illustrative case of an H atom bound to a flat face of the grain (see Fig. 2), the atom is bound to four surface-binding partners. The atom may also be seen to have four possible diffusion pathways, corresponding to diffusion along the surface {\em between pairs of binding partners}. While in theory there is an infinite number of paths that a diffusing atom may take, these four constitute those that pass through a saddle-point in the surface potential: they are, by far, the most probable paths. For each saddle-point pair, a diffusion barrier is defined equal to the total binding energy, {\em minus} the interaction potentials of the diffusing particle with each of the pair of atoms through which the saddle-point would pass. In other words, diffusion requires the breaking of two of the four bonds, while the other two remain intact, even after the hop is finished (Fig. 2). In a different case where, for example, the atom were bound to only 3 binding partners, only 1 bond would need to be broken. This approach is consistent with typical values of the ratio of diffusion barriers to binding energies, estimates of which can range from $\sim$0.3 -- 0.8 in current models (see e.g. Garrod \& Pauly 2011). Such values may be achieved through surface binding to different numbers of binding partners; the minimum of 3 partners (of equal interaction potential) would produce a value $E_{\mathrm{diff}}:E_{\mathrm{des}}=1/3$, while values closer to unity may be attained for binding to highly irregular (rough) surfaces.

The calculated diffusion barriers are stored and used to determine the selection of a particle and a thermal process as described in sections 2 and 2.3. If diffusion is selected, the next step is to determine the final position of the particle after diffusion through the selected saddle point.

In order to determine this position, the diffusion process is treated as a {\em rotation} around the saddle-point pair (with which the bonds remain intact following diffusion). The particle is rotated at a fixed radius around the axis joining the saddle-point pair, until it is close enough to be considered ``bonded'' to the atoms on the other side of the saddle-point.

At this stage, the position of the particle is optimized iteratively, to find the deepest part of the combined interaction potentials of the particles to which it is now bound. Each of these interaction potentials is treated as a Lennard-Jones 6--12 potential:
\begin{equation}
V_{LJ}=\, \epsilon \, \left[\left(\frac{\sigma}{s}\right)^{12} - 2 \, \left(\frac{\sigma}{s}\right)^{6}\right]
\end{equation}
where $s$ is the current center-to-center separation, $\epsilon$ is the depth of the potential, and $\sigma$ is the separation at which the potential minimum is reached, which is set to 3.2 \AA \, for the present model. (This expression is used wherever distance-dependent pair-wise potentials are required in the model; for example, the binding positions of newly-accreted particles are also calculated iteratively using this expression.) The derivatives of the potentials are used to calculate the net force vector resulting from the combined field at the current position. The position of the particle is advanced incrementally according to the force vector, to yield the position of the potential minimum to a tolerance of 0.005 \AA.

After the position is optimized in this way, the new binding energy is, again, taken simply as the sum of the strengths of the interaction potentials, $\epsilon$, with all the species to which the particle is bound (i.e. $0.9 \sigma < s < 1.1 \sigma$); new diffusion barriers for all {\em viable} saddle-point pairs are also calculated at this time.

The viability of saddle-point pairs is defined in the model as those pairs for which a rotation would be unhindered by other particles to which the diffusing species is {\em already} bound. For example, in the case shown in Fig. 2, while there are four pairs that would allow rotation (and thus diffusion), a pairing of {\em diagonally-arranged} particles from this group (Fig. 2c) is disallowed, because rotation/diffusion around the axis between the diagonal particles would be obstructed in either direction by another binding partner.

All rotation-pairings of particles to which the species in question is bound (i.e. $N(N-1)/2$ pairings, for $N$ bonds) are initially allowed. Each pairing is then assessed according to the following test: a plane is defined by the initial positions of the diffusing particle and each of the particles of the rotation pairing. If the {\em other} particles to which the surface species is bound are either {\em all above} or {\em all below} the plane, then diffusion is allowed (in a direction {\em away} from those particles). Otherwise, diffusion is disallowed, as a collision would result (as for the diagonal pairing shown in Fig. 2c). Typically, only pairings between adjacent (although not necessarily {\em bonded}) particles are allowed by this test, although irregular surface morphologies may allow other cases to be permitted.

In this way, the possible diffusion paths on an arbitrary surface may be determined. Species for which all rotation pairs are forbidden are considered ``boxed in'', and are thence also prohibited from desorption, until such time as this condition changes due to diffusion or evaporation of other particles.

It should be noted that, following accretion, desorption, or diffusion, the diffusion and desorption properties of all affected species -- such as binding partners both prior to and following a hop -- are re-calculated, and the rates and diffusion viabilities are re-assessed.

The rotation itself should not be considered an accurate description of the {\em motion} of the particle, but rather a convenient method for finding the potential minimum to which the particle hops, via a potential saddle-point. Actual diffusion paths for arbitrary binding structures would require intensive molecular dynamics or quantum chemical calculations.

In the case given in the example (Fig. 2), the diffusion barrier associated with each of the four possible hops is the same. However, if the particle were bound to different chemical species, differences in the magnitudes of the interaction potentials would lead to differences in the probabilities of diffusion along each path. On an irregular surface, the particle could be bound to 3 or more other particles; the {\em number} of binding partners as well as their chemical species is therefore important to the energy barrier to diffusion (i.e. the energy required to break all but two of the bonds).

The model is also optimized such that previous hops of the currently-active particle are stored in memory, so long as no other particles diffuse, desorb or accrete in the meantime. This decreases run times by several orders of magnitude; the calculations are often dominated by the diffusion of a single weakly-bound atom (until it finds a stronger binding site).

\subsubsection{Reaction}

In the case where either a diffusion or accretion event leads to a particle being bonded to a species with which it is allowed to react, reaction is assumed to occur immediately. The reaction product inherits the position of the {\em stationary} reactant, but the other physical characteristics of the new particle are re-calculated. Both the so-called ``Langmuir-Hinshelwood'' (diffusive) and ``Eley-Rideal'' (prompt) surface-reaction mechanisms are treated automatically in this model.

The current network includes the species H, O, H$_2$, O$_2$, OH, H$_2$O and H$_2$O$_2$, commonly referred to as the ``water system''. No activation energy barrier-mediated reactions are currently considered in the network; see Table 2.

\subsection{Other processes}

Many models include a number of other grain-surface processes not included here, such as photo-desorption and photo-dissociation. The latter is often suggested as an important mechanism for the formation of complex organic molecules in the interstellar medium (e.g. Garrod \& Herbst 2006). All such processes are expected to be included in future versions of the model, but are omitted here so that the basic approach may be tested. 

The inclusion in the model of photo-induced desorption based on estimated or measured rates (\"{O}berg et al. 2009) would be technically trivial. Photo-dissociation would be somewhat more complex, as it would require the positioning of two surface products that originated from a single molecule. 

Consideration of the formation of complex organic molecules using the new model is anticipated, but will depend on the adoption of a significantly larger reaction network that includes CO and its related products. Furthermore, only {\em surface} processes are considered in the current model; photo-processing of the sub-surface ice-mantle material would involve a considerable increase in technical complexity.

\section{Results}

Figure 1 shows a simulation of grain-surface chemistry using the method described above. The simulation uses a gas density of $n_{\mathrm{H}} = 2 \times 10^{5}$ cm$^{-3}$, and dust and gas temperatures of 10 K. The images show accurately the positions of each simulated particle; however, the size and appearance of each particle in the image is chosen purely for ease of visual identification. As described above, every atom and molecule in the present model is treated as a uniform sphere of radius 1.6 \AA; no inference as to the polar orientation of molecules should be drawn. The images are constructed using the free-ware ray-tracing software {\em POV-ray}.\footnotemark

\footnotetext{www.povray.org}

Panel (b) shows an instance wherein 2 oxygen atoms and 1 hydrogen atom have been accreted from the gas phase. All are free to diffuse, but the hydrogen atom is significantly more mobile than the oxygen atoms. Following several thermal hops, the hydrogen atom meets one of the oxygen atoms, with which it reacts to form OH. The further accretion of an oxygen and a hydrogen atom from the gas phase results in the formation of O$_2$ and H$_2$O. A video of this sequence may be found in the online version of the journal.

Figure 3 shows another simulation with a gas density of $n_{\mathrm{H}} = 2 \times 10^{5}$ cm$^{-3}$, extended for a much longer period; 1000 water molecules are formed over 350 yr. (An identical simulation is shown in Figs.\,\ref{f_surface}d and \ref{f_cross}d, extended to 200,000 water molecules, or 8,471 yr).

The resultant ice mantle is composed primarily of water molecules, as expected. Some molecular oxygen (O$_2$) and molecular hydrogen (H$_2$) are also present; panel (b) shows an H$_2$ molecule a little below the center of the image. To the left of this may be seen both an O$_2$ molecule and a hydrogen peroxide molecule (H$_2$O$_2$). H$_2$ and H$_2$O are in fact formed in approximately equal quantities, but the molecular hydrogen evaporates rapidly unless trapped in a strong binding site, typically one with $>$4 binding partners. Panel (c) shows that the H$_2$ indicated in panel (b) has left that binding site.

In rate equation-based models, the formation of any significant quantity of H$_2$O$_2$ by the direct addition of two OH radicals at low temperature would be unlikely, due to the low diffusion rate of OH. Such methods treat only the direct diffusion of each stated reactant. However, because the model presented here considers the explicit positions of all particles at all times, H$_2$O$_2$ may instead be formed by the diffusion of an O atom to a position that is in contact with an OH radical, followed by the surface diffusion of an H atom onto the oxygen atom. This results in an immediate two-stage reaction: H + O + OH $\rightarrow$ OH + OH $\rightarrow$ H$_2$O$_2$. The occurrence of this process is dependent only on the relatively fast diffusion of hydrogen and oxygen, and not OH. 

Such effects, which are essentially three-body surface reactions, are not generally taken into account in rate-based models. However, a similar process -- for the surface reaction of H + O + CO -- was considered by Garrod \& Pauly (2011), who found this mechanism to be the primary formation route for CO$_2$ ice at low temperatures, due to the immobility of CO and OH. Because the new model takes explicit account of physical positions and structure, all such multi-stage processes are included by default.

The formation of the ice mantle shown in Fig. 3 is not uniform; panels (b) and (c) in particular show the formation of a ``hole'' in the ice, corresponding to a region on the grain whose geometry results in weaker surface potential minima. In the center of the hole shown in panel (c) is a single dust-grain atom, around which six other dust-grain atoms reside. This arrangement produces six contiguous potential minima that allow bonding to only 3 dust-grain atoms, rather than the typical 4. Moderately mobile species that are momentarily bound in these weak sites rapidly diffuse out again. Thus, the geometry of the dust grain alone may be seen to influence the consequent ice-mantle structure. There exist 8 such regions on the grain used in these models, where similar behavior is found. In this particular model, these holes eventually close over, as molecules build up at the edges of the holes, presenting positions of stronger binding within. Under other conditions, such holes may progress to form pores within the ice structure (see section 3.1.1).

\subsection{Density-dependent models}

In order to investigate the formation of structure in the ices, simulations have been run to produce ice mantles several times thicker than the diameter of the underlying grain. Four sets of models have been produced, corresponding to gas densities in the range $2 \times 10^{4}$ -- $2 \times 10^{7}$ cm$^{-3}$. For each density, the model has been run three times, adopting a different initial random number seed. This is done to demonstrate that the behavior of the models is a general feature and not dependent on the specific random numbers used in each run. All the simulations are run until 200,000 water molecules have been formed on the grain. As before, dust and gas temperatures are held at 10 K and the gas-phase fractional abundances of H and O atoms are fixed at $2 \times 10^{-4}$.

In addition, a single model has been run in which water molecules are accreted directly from the gas phase, with no accretion of H and O atoms. An arbitrary gas density of $2 \times 10^{7}$ cm$^{-3}$ is used; however, for the consideration of ice structure, the precise rate of accretion is unimportant, due to the lack of surface chemistry and the neglible rate of water diffusion caused by the low temperature.

\subsubsection{Structure and porosity}

Figure 4 shows the final state of each density model (run 1 is shown in each case), including pure accretion (panel a). Hydrogen molecules (H$_2$) are highlighted in blue to aid the eye.

A number of features are immediately apparent; firstly, the large- and small-scale structure of the ice is strongly dependent on the gas density. The most irregular structure results from the higher-density simulations, culminating in the accretion-only model -- in this case, water molecules remain wherever they land as the result of accretion. In the simulations with active chemistry, higher densities result in more rapid hydrogenation of surface oxygen atoms, before they have the opportunity to find alternative, stronger binding sites.

The higher-density ice mantles exhibit a highly porous, or ``creviced'' structure. Panels (b) and (c) in particular show a ``cauliflower-like'' structure, wherein the crevices act to isolate large-scale nodules of ice. In the lower density simulations, the nodules become broader and the crevices less deep. The lowest-density model indeed shows a fairly smooth ice structure with no deep crevices or voids. 

In the accretion-only simulation, the ice appears irregular on all scales. The small-scale porous structure is largely the result of the lack of surface diffusion, such that new species are unable to move even a short distance into the stronger binding sites that are provided by areas of extreme small-scale curvature of the ice surface. However, the large-scale irregularity of the ice is caused by the combination of fully-randomized trajectories with a three-dimensional grain; the growth of randomly-produced surface irregularities, or ``bulges'', is amplified by their protrusion into the accretion field, picking up more material from a greater solid angle, while blocking out those trajectories for other regions of the grain/ice surface. Such effects are still prevalent in the active-chemistry models -- panel (c) shows an ice mantle whose underlying dust grain is centered in the image; a large protrusion may be seen on the right. While a moderate degree of surface diffusion may act to close over the small-scale pores or crevices in the ice, atoms of oxygen are required to diffuse a significant distance over the ice surface to counteract the large-scale irregularities. Consequently, while smooth on the small scale, the lowest-density simulation (panel e) nevertheless shows an overall irregular shape.

Figure 5 shows cross sections of the same final ice mantles. Each cross section passes through the dust-grain center at an arbitrary angle, showing molecules that sit within a ``slice'' $3\sigma$ deep -- this allows space for up to two molecules along a line of sight. Videos of panels (a), (b) and (e) are available in the online version of the journal, showing cross sections over 360 degrees.

The detail of the porosity within the ice mantles is clearly visible in Figure 5. Panel (a) shows significant and fairly uniform porosity, with pores or crevices that pass deep within the mantle almost down to the dust grain itself. The highest-density simulation with active chemistry is not as extreme, but also shows very deep pores/crevices in places, and exhibits some closed pores. The degree of porosity falls away at lower gas densities, until, for $n_{\mathrm{H}}=2 \times 10^{4}$ cm$^{-3}$, there are essentially no open pores and the ice is extremely compact.

One of the most striking features of Fig. 5 is the arrangement of hydrogen molecules (H$_2$). Shown in blue in Figs. 4 and 5 for ease of identification, H$_2$ is almost completely segregated from the water molecules, especially in the higher-density simulations. Furthermore, the H$_2$ molecules are arranged in ``veins'' that are similar in appearance to the empty porous structures. As may be seen three-dimensionally in Fig. 4, the H$_2$ indeed fills in crevices between larger structures. The veins of H$_2$ are typically no more than around 5 molecules across (commensurate micro-porous structure), while many of the unfilled pores are significantly larger.

Following its formation on the grains, an H$_2$ molecule may diffuse around the surface. If it finds a weak binding site, it may evaporate entirely, but if it finds a strong binding site -- such as may be found in a pore, where multiple binding potentials converge -- then it may remain bound in place. The addition of further hydrogen molecules could then contain it, and allow a pore to fill up with hydrogen. The presence of larger, unfilled pores in the higher-density simulations suggests that these pores have an insufficient degree of inward curvature to produce binding potentials great enough to retain H$_2$ for long enough before the deposition of new material can lock it into place.

Figure \ref{bar_coord} shows a bar chart of the number of surface bonds of each molecule (of any species) within the ice mantle, for each of the simulations shown in Figs. 4 and 5. The accretion-only model shows a very different distribution from the active-chemistry models. Approximately 35\% of all molecules in this simulation experience only 3 -- 5 bonds, making them (typically) surface molecules. Conversely, only around 6\% of molecules in even the highest-density active-chemistry simulation are so weakly bound, and the number of species with only 3 bonds is negligible. However, this does not imply that potential minima that would allow only 3 bonds do not exist -- only that they are unoccupied. Pores, and the associated H$_2$ veins, may begin to form in such positions, with ice structure building up around them.

The porosity visible in Figs. 4 and 5 is reflected in the bonding distributions of Fig. 6, with the most compact ice (produced by the lowest-density model) showing the greatest bias toward 10 -- 12 bonds, which are necessarily bulk ice molecules. Such distributions may therefore be useful as a measure of porosity in these simulations.

Figure 7 shows the fraction of bonds shared with an H$_2$ molecule, averaged over all H$_2$ molecules. The highest density model shows the greatest degree of clustering of H$_2$ molecules. Based on a purely statistical comparison of abundances, the expected value would be around 13 -- 14\% for this density (based on values in Table 3). For the lowest density model, the statistical expectation would be $\sim$8\%. In each case, the actual fractional bonding of H$_2$ to H$_2$ is close to 3 times higher than the statistical expectation. It should also be borne in mind that the binding potentials used for H$_2$ (Table 2) are the same for all binding partners. The H$_2$ clustering is thus the result of the structural arrangement of the ice mantle to produce pores or concentrations of strong binding sites in which H$_2$ molecules may become trapped, rather than a preference for H$_2$ molecules to bind to each other, per se.

\subsubsection{Chemistry}

Figure 8 shows time-dependent grain-surface chemical abundances for each of the stable molecules included in the model, for all four gas-density values. Results for all three runs at each density are plotted in the same panel. The end-time of each model corresponds to the time at which 200,000 water molecules have been formed; the precise value varies according to the gas density and, to some degree, according to random variation between runs at the same density. For O$_2$ and H$_2$O$_2$, which have generally low abundances, there is some random fluctuation apparent at early times, and there are small systematic differences between runs of the same density. However, as a proportion of the total quantities, these variations fall away as the ice mantles grow.

Table 3 shows the final values for each model run, as well as the total amount of H$_2$ formed, much of which evaporates back into the gas phase. H$_2$O$_2$ shows the largest variation between same-density runs, of around 7-8\%, but variations for the other species are typically 1-2\%. The elevated value for H$_2$O$_2$ may be due to its more intricate formation mechanism in this model (see section 3). For reference, Table 3 also indicates end-times of the models, $t_{\mathrm{f}}$, and the greatest radial distance of any ice-mantle particle from the centroid of the grain, $r_{\mathrm{max}}$, achieved by the end of each run.

Abundances of species other than H$_2$O are seen to be somewhat lower for the lower density models, including H$_2$. However, the total amount of H$_2$ formed (see Table 3) is much less affected than the quantity that is stored on the grain at each gas density. Accreted atomic hydrogen is typically able to remain on the grain for long enough for another reactive species to accrete and react with it, for all density models. In the case where the product is H$_2$, it is more likely to evaporate for the lower density models, which have a more regular surface that provides fewer strong-binding positions or pores. In each of the simulations, the amount of H$_2$ formed is only a small amount larger than the total water formed.

\subsection{Computational considerations}

The running of the new model over various density conditions allows an assessment of the computational efficiency to be made. Each simulation was run with an {\em Intel Xeon X5482} cpu, using a single thread.

The run-time for all of the three higher-density simulations ($2 \times 10^{5}$, $2 \times 10^{6}$, and $2 \times 10^{7}$ cm$^{-3}$) is approximately 90 -- 95 hours, depending on which random-number seed run is considered. Only the lowest-density simulations show an apparent density dependence, taking $\sim$180 hours to run.

The majority of cpu-time is used up on the calculation of diffusion pathways and on the assignment of position and pairing information associated with each surface hop. Because the model keeps diffusion paths in memory, the most intensive calculations are typically done only once per particle, providing the positions of surface potential minima for an individual, mobile particle for large numbers of subsequent, identical hops, prior to reaction, desorption or trapping in a strong binding site. 

It appears that the larger drain on cpu-time in the low-density case is caused by the relative smoothness of the ice surface. In this case, a diffusing particle (typically atomic hydrogen) on the grain has a smaller probability of becoming trapped in a strong binding site (thus ceasing its diffusion), while the probability of another particle being accreted onto the grain, and subsequently reacting with the first particle, is smaller (due to lower gas density). This results in a longer sequence of diffusion before a reaction occurs.

It is to be expected that the use of larger grains would ameliorate this effect, as the current model dust grain is extremely small, providing low overall rates of accretion. Conversely, higher temperatures would most likely make it worse, due to the resulting faster diffusion, although desorption rates would also increase.

The inclusion of a broader network of surface species is expected to make the surface more heterogeneous, such that the smoothness of the surface potential would be reduced, minimizing the possibility that surface species may spend long periods of cpu-time without being trapped in a strong binding site.

\section{Discussion}

\subsection{Porosity}

This model allows porous structures within the ice mantle to form naturally, as the result of chemical and physical conditions. However, the consideration of porosity in simulations of interstellar dust-grain chemistry is not new; Perets \& Biham (2006) used a rate-based model to study the effects of porosity within the dust-grain structure itself. The models indicated that porosity could increase the efficiency of H$_2$ formation over a wider range of temperatures than would otherwise be possible. Taquet et al. (2012) extended the treatment of Perets et al. to a full gas-grain chemical model that included the formation of a porous ice mantle, using pre-defined values for the number and sizes of pores. They found only moderate increases in the production of certain key ice-mantle species.

Cuppen \& Herbst (2007) used a Monte Carlo technique to treat the formation of ice on a surface with periodic boundary conditions, using a fixed lattice for the positions of particles on the grain or within the ice mantle. Depending on the physical conditions, ice structures of varying compactness and porosity were produced, including some with tower-like structures of apparently single-molecule thickness. The models that show the most extreme porosity, however, were obtained assuming temperatures rather higher than the 10 K used here.

In the present model, only a limited reaction set is considered, which does not include any activation barrier-mediated reactions. Nevertheless, the basic reactions that allow the formation of the most important species, such as H$_2$ and H$_2$O, are indeed present. The use of an off-lattice approach in this model requires that the binding of atoms or molecules to the surface involve at least 3 other binding partners. Thus, the ``sky-scraper'' structures produced by the models of Cuppen \& Herbst are not found here, both because binding to a single partner cannot occur and because the direction of the binding is not pre-determined, but depends on the arrangement of the binding partners. 

Perhaps most importantly, the off-lattice approach of the present model allows the effects of the curvature of the surface to be automatically taken into account. Inward curvature, such as may be found within the pores, produces stronger binding conditions resulting from the ability to bind with a greater number of partners; likewise, the outward curvature found on the bare grain, and on the ``nodules'' or ``bulges'' that form on the ice, tend to minimize the number of binding partners. These curvature effects drive mobile particles into the pores as the pores are forming, so that they become filled with H$_2$, producing veins. Larger pores -- which are formed in the higher-density models, and which experience less extreme curvature -- tend not to fill up in this way.

The kind of micro-pores investigated by Taquet et al. (2012), which appear to be comparable in size to the H$_2$ veins that form here, may not therefore be long-lived enough to affect the chemistry significantly. The present model indicates that such pores would be filled up, although others may develop in their place. In general, however, the low density model (which perhaps provides the most appropriate comparison) tends to predict a compact, albeit segregated, ice structure.

The gas-phase chemical abundances used in the model presented here represent dense-cloud conditions, under which most hydrogen takes the form of H$_2$, resulting in similar H and O gas-phase abundances. However, higher values for H are entirely plausible, which would result in greater H fluxes onto the grains, and therefore a shorter time period before surface oxygen atoms could find alternative binding sites. It is therefore likely that a cloud with large quantities of H not yet converted to H$_2$ should produce more porous dust-grain ices. It is also probable that, other model quantities being equal, the smallest of these pores would rapidly fill up.

The consideration here of a model that involves the direct accretion of water onto the grain allows a basic comparison with laboratory amorphous solid water (ASW) ices, which are also formed by the deposition of water molecules directly onto a cold surface. It is well known that the angle of deposition has a strong influence on the degree of porosity in the ice (e.g. Kimmel et al. 2001; Raut et al. 2007). The present model does not investigate the effect of deposition angle directly, but uses a completely randomized field of accreting particles. The result is an extremely porous ice, which is qualitatively different from the forms produced by active grain-surface chemistry, because of the minimal diffusion of the water molecules following accretion. In the active-chemistry models, the slow formation of the ice from its constituent atoms allows voids and strong binding sites to be filled more effectively, so that the lowest-density simulations show an absence of pores, and a relatively small quantity of H$_2$ veins as compared with higher gas-density (i.e. faster-accretion) models. 

It therefore seems appropriate to suggest that interstellar ice analogs formed in the laboratory may be significantly more porous than actual interstellar ices. This would imply that effects observed in laboratory ices, relating to the absorption of other molecules into pores and their subsequent release at higher temperatures (e.g. Collings et al. 2004), may be of less importance in interstellar ice mantles. In mitigation, the present model does not include any mechanism for super-thermal surface diffusion of the accreted particle, or indeed of newly-formed surface molecules. The acceleration of the accreting particle into a surface potential well, or the release of chemical energy from the surface formation of a molecule, could provide sufficient energy for diffusion. This could allow an otherwise immobile particle to find a nearby binding site with stronger binding properties. However, in view of the experimental evidence for significant porosities in laboratory ASW ice (e.g. Westley et al. 1998, and references therein), the importance of this effect may be small.

\subsection{Chemistry and kinetics}

As shown in Table 3, for decreasing density the H$_2$:H$_2$O production ratio approaches 1:1. This may be explained through a simple analysis: Taking the 1:1 gas-phase abundance ratio for H:O used in the present model, the ratio of accretion rates is 4:1, due to the 16 times higher mass of oxygen, as entered into Eq. (1). Under conditions where atomic-hydrogen mobility dominates all surface processes, and accretion dominates evaporation, this rate should indeed produce a 1:1 formation rate for H$_2$:H$_2$O. (For every 1 oxygen that accretes, 4 H-atoms will accrete; 2 are used up, as they rapidly find the O/OH on the surface, to form water. The other two will -- on average -- meet each other and form H$_2$.) This analysis ignores O$_2$ and H$_2$O$_2$ formation, each of which would otherwise remove 4 and 2 more hydrogen atoms, respectively, if their oxygen atoms had instead formed surface water. This difference accounts for around 40\% of the H$_2$ excess over H$_2$O in the low-density models.

It may be seen, therefore, that for the lowest density simulations in particular, the production of H$_2$ is very close to what may be called a ``smooth-grain'' limit, in which hydrogen-atom mobility is not significantly affected by the presence of strong binding sites. 

With increasing density, there are more strong binding sites (due to the more irregular ice structure), while the waiting period for the accretion of a new particle is also shorter. These effects heighten the probability that an accreted hydrogen atom will wait for another H to react with it, rather than to diffuse to find a free oxygen atom elsewhere on the dust/ice surface, thus producing a greater proportion of molecular hydrogen.

For gas densities less than $\sim$$2 \times 10^{4}$ cm$^{-3}$, one should expect that the ice surface would be similarly smooth, and perhaps yet more spherical, while the waiting period for accretion would be greater, all of which should produce H$_2$:H$_2$O ratios even closer to unity. Under such circumstances, the use of rate-based models that employ generalized binding and diffusion characteristics should not produce strongly divergent results for H$_2$ production from those obtained here, providing appropriate characteristic values were chosen. Such an approach would also require that the stochastic behavior of the chemistry be treated adequately (e.g. the use of the modified rate equations of Garrod 2008 and Garrod et al. 2009), as the scenario described clearly falls into the case where standard rate equations should fail. It is unclear, however, whether such an approach could accurately reproduce the quantity of the resultant H$_2$ that is retained on the grain surface, which is still determined by heterogeneous structural considerations.

\subsection{Physical conditions}

The choice of physical conditions investigated in the present study has been limited to the gas density, which controls the rate of accretion onto the grain. The gas density demonstrates a clear effect on the resultant ice structure, and thence on the retention of H$_2$ molecules on the grain. The relative formation rates of O$_2$ and H$_2$O$_2$ are also seen to be affected by the density variation, due to the competition between chemical reactions whose rates are dependent on surface diffusion.

Each of the effects mentioned above concern the interplay between accretion, diffusion and desorption. The gas density is thus not the only parameter which may affect the chemistry and structure of the ice formed on the grain. Future work will address the importance of dust temperature; however, one may predict that temperatures higher than the 10 K used here would produce more diffusion of oxygen atoms, meaning more compact ices. Likewise, lower temperatures would result in more porosity, and greater retention of H$_2$ within those structures. 

The present, preliminary model uses a very small grain, of radius 16 \AA, as compared to a canonical value of 0.1 $\mu$m (=1000 \AA). However, the final ice-mantle radii are significantly larger, at around 150 -- 190 \AA. Future models will investigate chemistry on larger grains. The application of the model to a porous dust grain, of explicitly defined structure, would also be quite possible.

\section{Conclusions}

The model presented in this paper is the first off-lattice Monte Carlo kinetic model of interstellar dust-grain chemistry. The model allows the full three-dimensional simulation of chemical kinetics and ice structure on a grain surface defined by the positions of its constituent atoms. The use of an off-lattice technique allows the precise positions, binding strengths and diffusion probabilities of grain-surface particles to be determined according to the interaction potentials with binding partners; particles are not fixed in a pre-determined lattice structure. The model also allows arbitrary morphologies and structures to be used for the underlying dust grain. This includes not only the local surface roughness of a particle, but the degree of grain porosity, as well as the choice of a spherical, spheroidal, irregular, or any other grain shape. The model opens up a large new parameter space that will be investigated in future, in conjunction with a more extensive chemical network.

The main conclusions of this preliminary study are summarized below:
\begin{enumerate}
\item The shape of the ice mantle is irregular, even for low gas-density models. 

\item The porosity of the ice is strongly dependent on the gas density, with higher densities producing greater porosity.

\item The small-scale curvature of local regions of the ice/grain surface has a strong effect on the chemistry and structure of the ice formed on top. Such considerations can only be treated using an off-lattice approach.

\item Inward curvature on small scales -- as found with micropores -- allows binding with a greater number of binding partners than on a relatively flat surface. This increases desorption and diffusion barriers, encouraging the trapping of mobile species in micropores and strong surface binding sites.

\item Micropores are formed in the ices, but are rapidly filled with H$_2$ molecules, forming veins of H$_2$. Thus, H$_2$ shows significant segregation from water molecules in all models.

\item Larger pores remain unfilled, as they do not provide sufficient small-scale curvature of the surface to significantly increase binding potentials.

\item For gas densities appropriate to dark interstellar clouds ($2 \times 10^{4}$ cm$^{-3}$), the ice mantle is fairly smooth and non-porous, albeit with veins of H$_2$ throughout.

\item Direct deposition of water molecules, such as is used to produce laboratory amorphous water ice, results in far greater porosity than is achieved when the ice is formed by surface reactions between accreted atoms.

\end{enumerate}

\acknowledgements

This work was partially funded by the NASA Astrophysics Theory Program, grant number NNX11AC38G. The author thanks the anonymous referee for helpful comments.

\begin{figure}
\begin{center}
\includegraphics[width=0.49\textwidth]{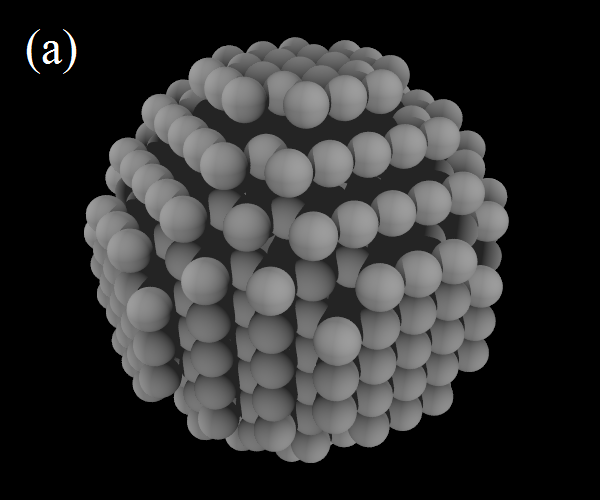}
\includegraphics[width=0.49\textwidth]{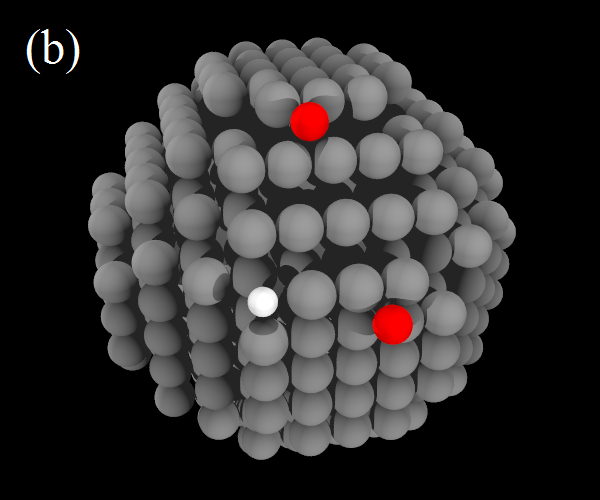}
\includegraphics[width=0.49\textwidth]{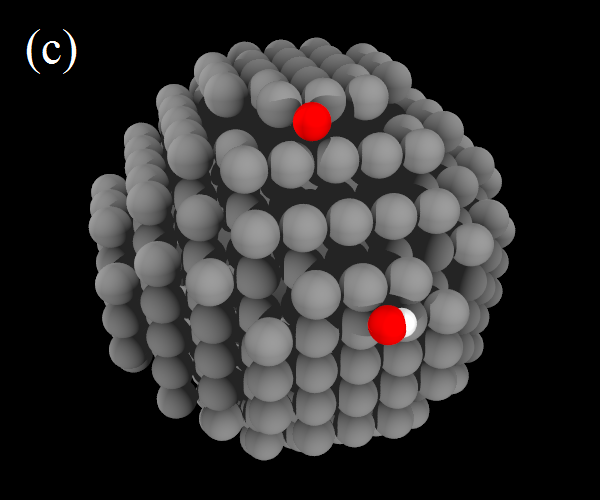}
\includegraphics[width=0.49\textwidth]{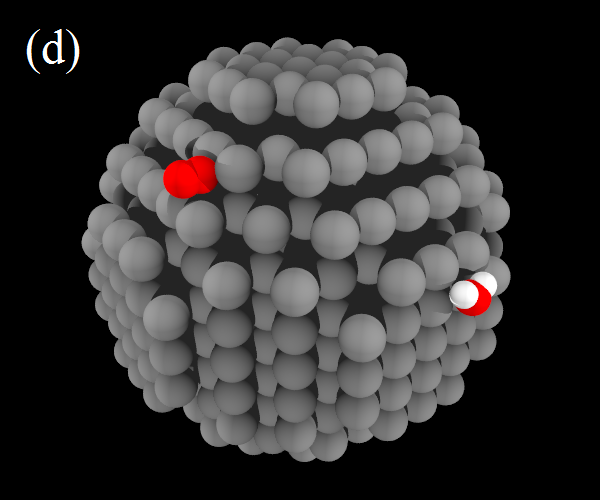}
\end{center}
\caption{{\bf (a)} Small, idealized grain used in this model, with simple cubic structure; {\bf (b)} simulation, with 2 oxygen (red) atoms and 1 hydrogen (white) atom accreted from gas phase and bound to surface; {\bf (c)} hydrogen atom has diffused across grain surface to meet an oxygen atom, forming OH; {\bf (d)} an oxygen atom has accreted, diffused, and reacted with another to form O$_2$, then another hydrogen atom has landed, to react with OH, forming H$_2$O. {\em A video of this sequence is availble in the online version of the journal, showing one thermal hop per frame.}}
\end{figure}

\begin{figure}
\begin{center}
\includegraphics[width=0.32\textwidth]{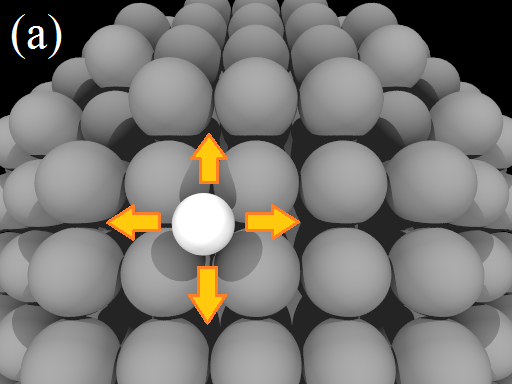}
\includegraphics[width=0.32\textwidth]{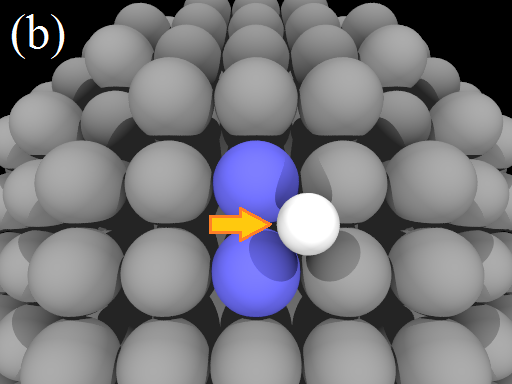}
\includegraphics[width=0.32\textwidth]{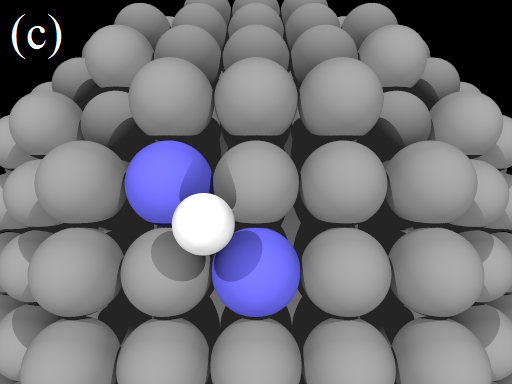}
\end{center}
\caption{{\bf (a)} The four possible diffusion pathways for an H atom bound to a flat face of the grain; {\bf (b)} After the diffusion event, the H atom remains bound to two of the original binding partners (the saddle-point pair, highlighted), as well as to two new ones; {\bf (c)} For this surface arrangement, rotation around the axis joining a diagonal pair (highlighted) is forbidden, because the diffusion path is obstructed in each direction by another binding partner.}
\end{figure}

\begin{figure}
\begin{center}
\includegraphics[width=0.32\textwidth]{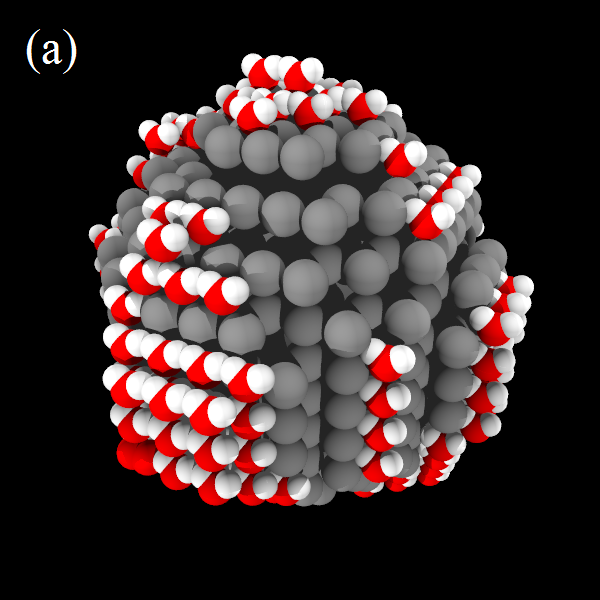}
\includegraphics[width=0.32\textwidth]{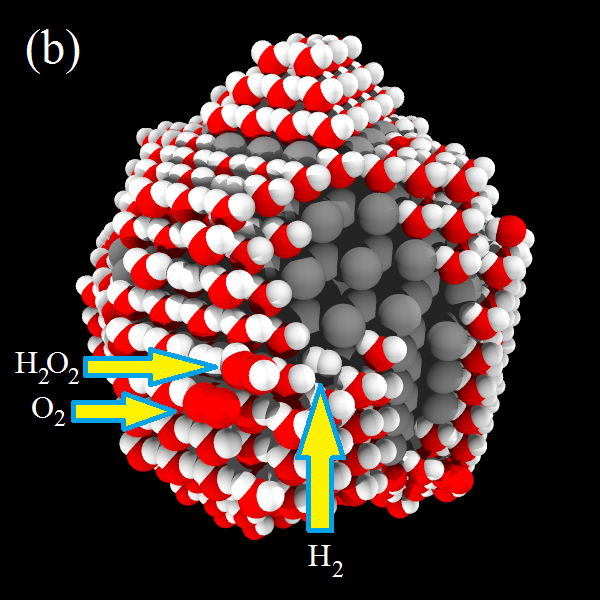}
\includegraphics[width=0.32\textwidth]{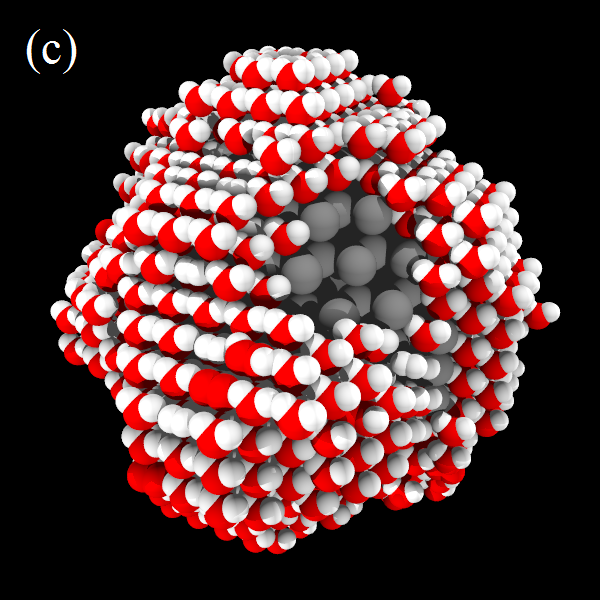}
\end{center}
\caption{Gradual build-up of an ice mantle on the grain surface, for a simulation with gas density $n_{\mathrm{H}} = 2 \times 10^{5}$ cm$^{-3}$. The ``hole'' in the ice structure is caused by the comparatively weak surface potential of the underlying grain surface in that region, resulting from its geometry. {\em A video of this simulation is availble in the online version of the journal, showing the formation of 1 water molecule per frame.}}
\end{figure}

\begin{deluxetable}{lrrrrrrrr}
\tabletypesize{\small}
\tablecaption{\label{tab-pot} Pair-wise interaction potentials, in units of K.}
\tablewidth{0pt}
\tablecolumns{9}
\tablehead{
               &  \colhead{Grain}  &  \colhead{H}  &  \colhead{ H$_2$}  &   \colhead{O}  &   \colhead{O$_2$}  &   \colhead{OH}  &   \colhead{H$_2$O}  &  \colhead{ H$_2$O$_2$} }
\startdata
H                  &  100  &  100  &    50  &  100  &  100  &  100  &  100  &  100  \\
H$_2$           &    50  &    50  &    50  &   50  &    50  &    50  &    50  &   50  \\
O                  &  200  &  100  &    50  &  200 &   200  &  200  &  200  &  200  \\
O$_2$           &  300  &  100  &    50  &  200 &   300  &  300  &  300  &  300  \\
OH                &  400  &  100  &    50  &  200 &   300  &  400  &  500  &  600  \\
H$_2$O         &  500  &  100  &    50  &  200 &   300  &  500  & 1000 & 1000 \\
H$_2$O$_2$  &  600  &  100  &    50  &  200 &   300  &  600  & 1000 & 1200 \\
\enddata
\end{deluxetable}

\begin{deluxetable}{llrl}
\tabletypesize{\small}
\tablecaption{\label{tab-reac} Surface reactions}
\tablewidth{0pt}
\tablecolumns{4}
\tablehead{
\colhead{\#} & \colhead{Reaction} }
\startdata
 1 & H        +  H         &      $\rightarrow$ & H$_2$              \\ 
 2 & O        +  O         &      $\rightarrow$ & O$_2$             \\ 
 3 & H        +  O          &     $\rightarrow$ & OH                  \\ 
 4 & H        +  OH         &    $\rightarrow$ & H$_2$O           \\ 
 5 & OH      + OH           &    $\rightarrow$& H$_2$O$_2$   \\ 

\enddata
\tablecomments{All reactions are presumed to proceed without an activation barrier.}
\end{deluxetable}

\begin{deluxetable}{rcccccccccccc}
\tabletypesize{\small}
\tablecaption{\label{tab-fin} Final time, maximum radius, and final number of grain-surface particles, by species, bound to the grain (and, in the case of H$_2$, the total formed on the grain), for each model run. All models were stopped after the formation of 200,000 water molecules.}
\tablewidth{0pt}
\tablecolumns{13}
\tablehead{
\colhead{$n_{\mathrm{H}}$}  &  \colhead{Run} & \colhead{} &  \colhead{$t_{\mathrm{f}}$} &   \colhead{$r_{\mathrm{max}}$} &     \colhead{} &  \colhead{H}  &  \colhead{O}  &   \colhead{OH}  &   \colhead{O$_2$}  & \colhead{H$_2$O$_2$}  &  \colhead{H$_2$}  & \colhead{H$_2$} \\
\colhead{(cm$^{-3}$)} & & & (yr) & \colhead{(\AA)} & & & & & & &  \colhead{(on grain)} & \colhead{(total)}
}
\startdata
$2 \times 10^{4}$ & 1 && $7.047 \times 10^{4}$ & 147.1 && 3 & 1 & 3 & 1,238 & 240 & 18,284 & 207,313 \\
                            & 2 && $6.916 \times 10^{4}$ & 152.6 && 6 & 3 & 3 & 1,241 & 261 & 18,399 & 206,096 \\
                            & 3 && $6.994 \times 10^{4}$ & 149.1 && 1 & 4 & 0 & 1,260 & 257 & 18,407 & 208,248 \\
\\
$2 \times 10^{5}$ & 1 && $8.471 \times 10^{3}$ & 153.9 && 6 & 7 & 8 & 1,599 & 304 & 23,011 & 209,973 \\
                            & 2 && $8.214 \times 10^{3}$ & 148.3 && 9 & 4 & 5 & 1,547 & 301 & 22,998 & 206,612 \\
                            & 3 && $8.175 \times 10^{3}$ & 160.0 && 7 & 6 & 8 & 1,523 & 325 & 23,494 & 208,044 \\
\\
$2 \times 10^{6}$ & 1 && $8.216 \times 10^{2}$ & 164.9 && 37 & 30 & 19 &  3,073 & 620 & 28,872 & 217,227 \\
                            & 2 && $8.292 \times 10^{2}$ & 176.0 && 27 & 23 & 19 &  2,974 & 618 & 29,216 & 217,463 \\
                            & 3 && $8.298 \times 10^{2}$ & 178.3 && 27 & 20 & 18 &  3,031 & 596 & 29,620 & 216,183 \\
\\
$2 \times 10^{7}$ & 1 && $6.687 \times 10^{1}$ & 191.9 && 72 & 42 & 52 &  5,308 & 1,321 & 32,824 & 225,258 \\
                            & 2 && $6.695 \times 10^{1}$ & 177.3 && 60 & 53 & 45 &  5,385 & 1,233 & 31,484 & 228,142 \\
                            & 3 && $6.978 \times 10^{1}$ & 183.6 && 81 & 42 & 42 &  5,381 & 1,281 & 33,319 & 227,244 \\
\enddata
\end{deluxetable}

\begin{figure*}
\center
\includegraphics[width=0.49\textwidth]{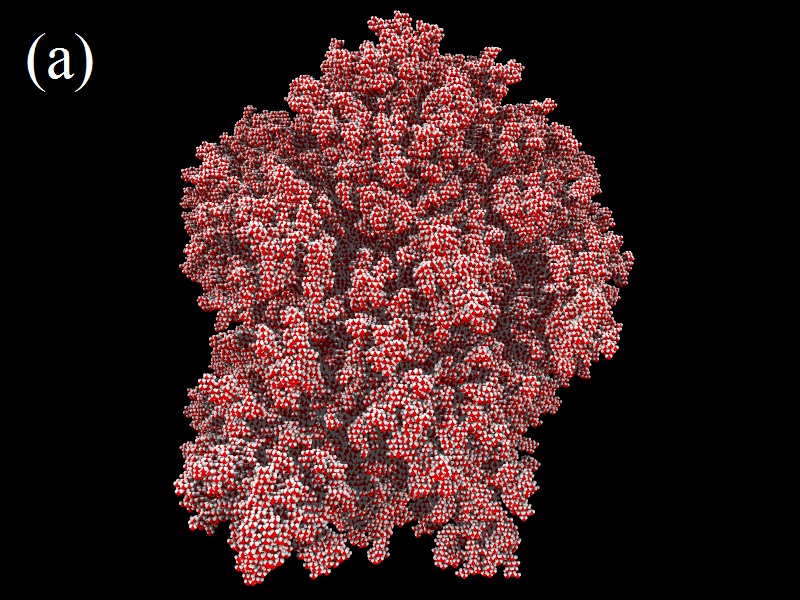}
\includegraphics[width=0.49\textwidth]{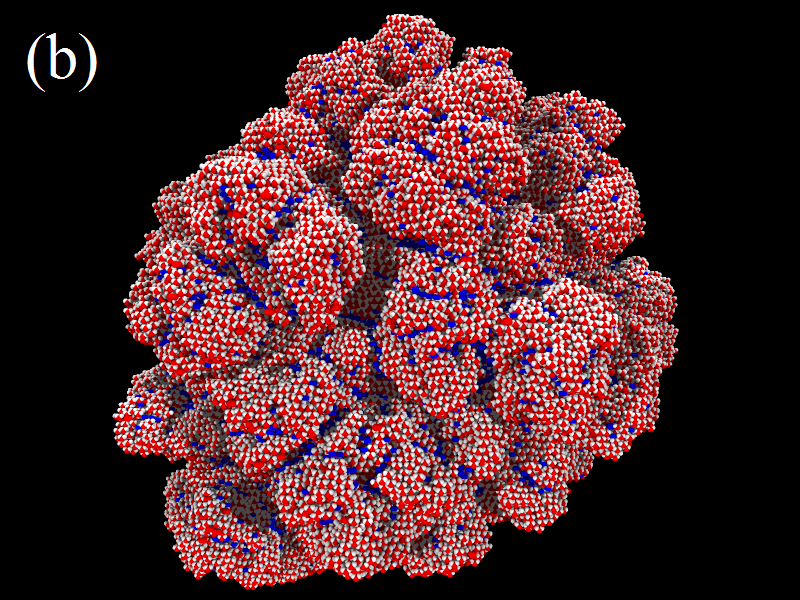}
\includegraphics[width=0.49\textwidth]{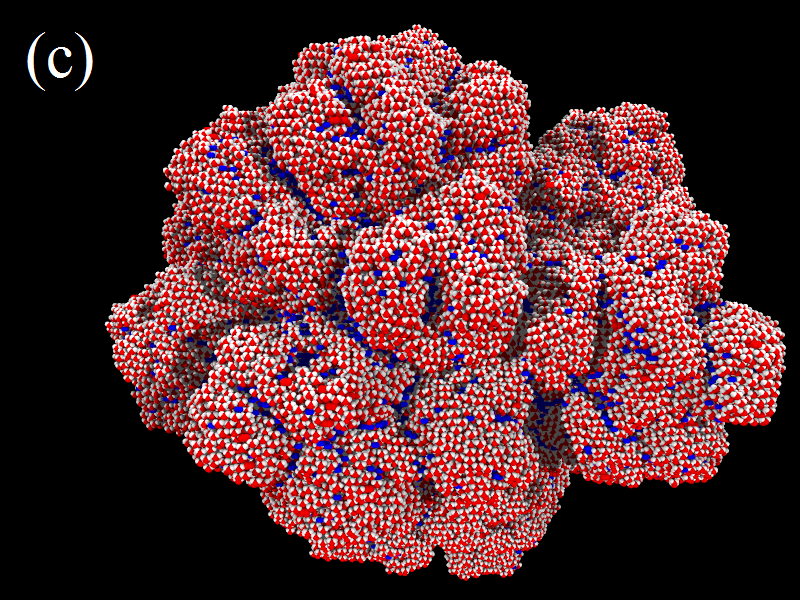}
\includegraphics[width=0.49\textwidth]{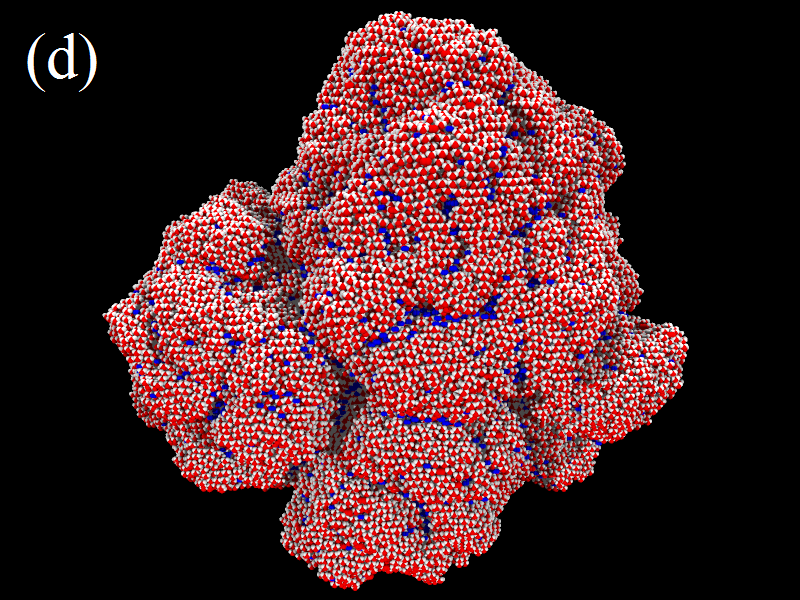}
\includegraphics[width=0.49\textwidth]{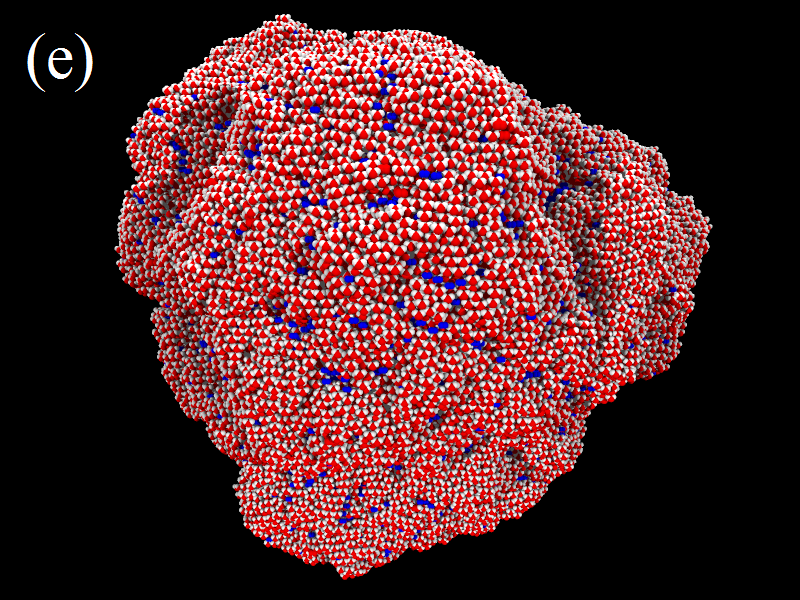}
\caption{\label{f_surface} Simulated dust-grain ice mantles formed at various gas densities, with each containing 200,000 water molecules, as well as other species; molecular hydrogen (H$_2$) is highlighted in blue to aid the eye. {\bf (a)} Mantle formed through direct accretion of H$_2$O, with no surface chemistry and arbitrary gas density; {\bf (b)} mantle formed through accretion and surface chemistry of H and O atoms, with a gas density of $n_{\mathrm{H}} = 2 \times 10^{7}$ cm$^{-3}$; {\bf (c)} $n_{\mathrm{H}} = 2 \times 10^{6}$ cm$^{-3}$ ; {\bf (d)} $n_{\mathrm{H}} = 2 \times 10^{5}$ cm$^{-3}$ ; {\bf (e)} $n_\mathrm{H}{} = 2 \times 10^{4}$ cm$^{-3}$. {\em Videos of panels (a), (b) and (e) may be found in the online version of the journal, showing the addition of 200 water molecules per frame.}}
\end{figure*}

\begin{figure*}
\center
\includegraphics[width=0.49\textwidth]{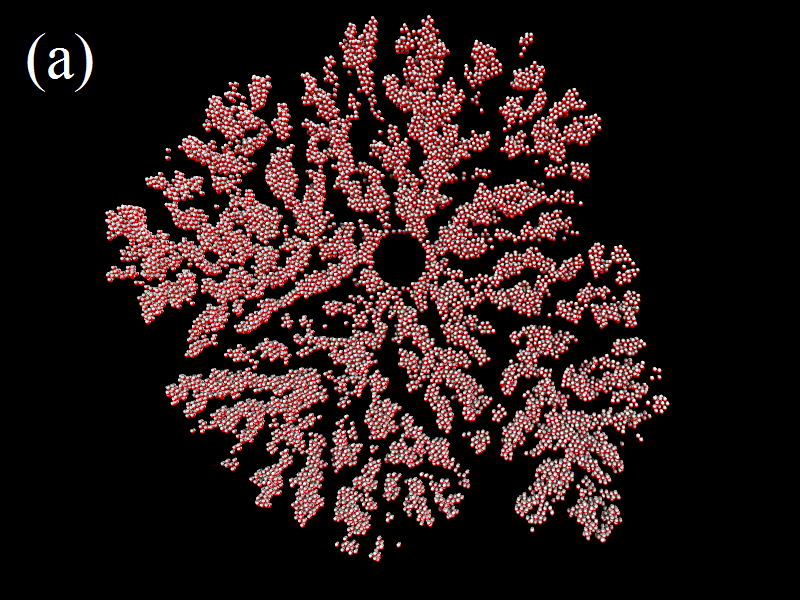}
\includegraphics[width=0.49\textwidth]{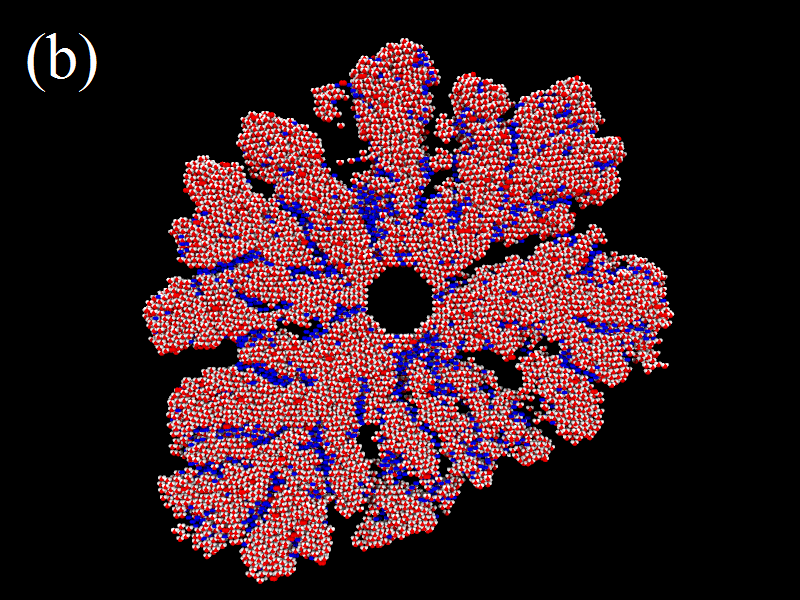}
\includegraphics[width=0.49\textwidth]{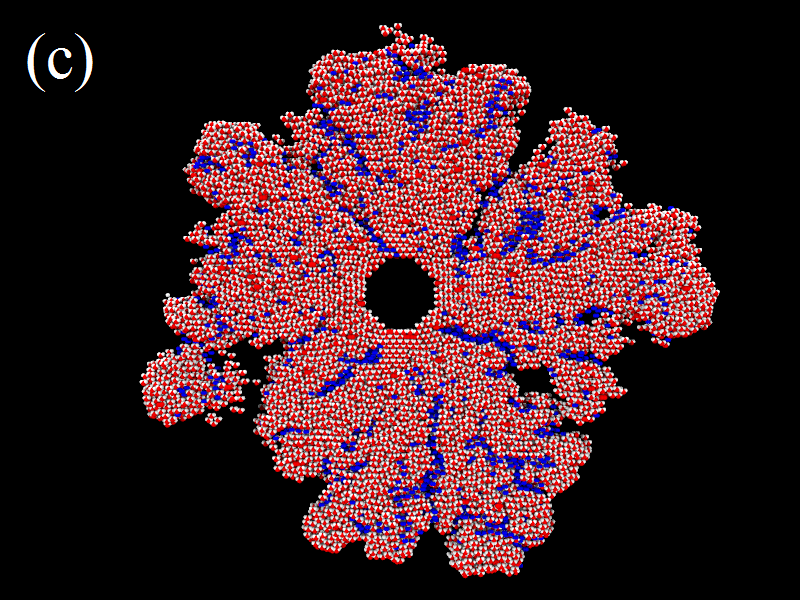}
\includegraphics[width=0.49\textwidth]{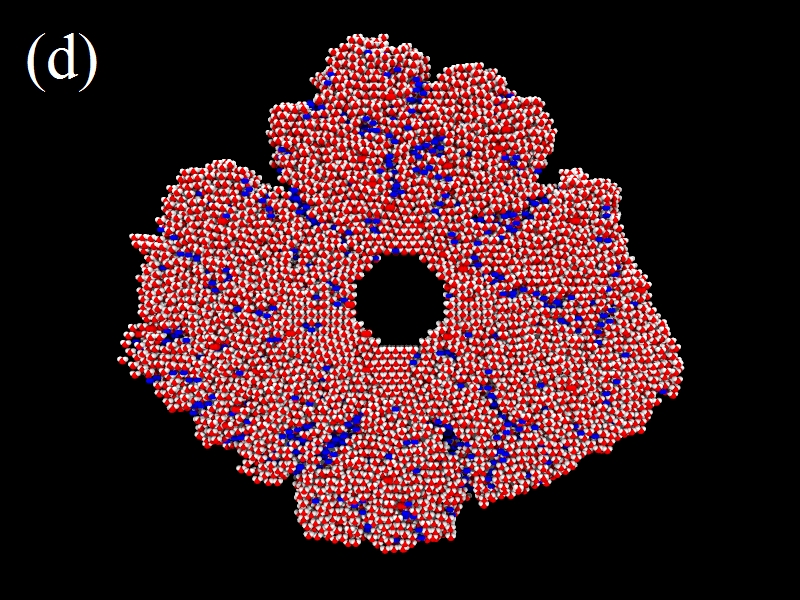}
\includegraphics[width=0.49\textwidth]{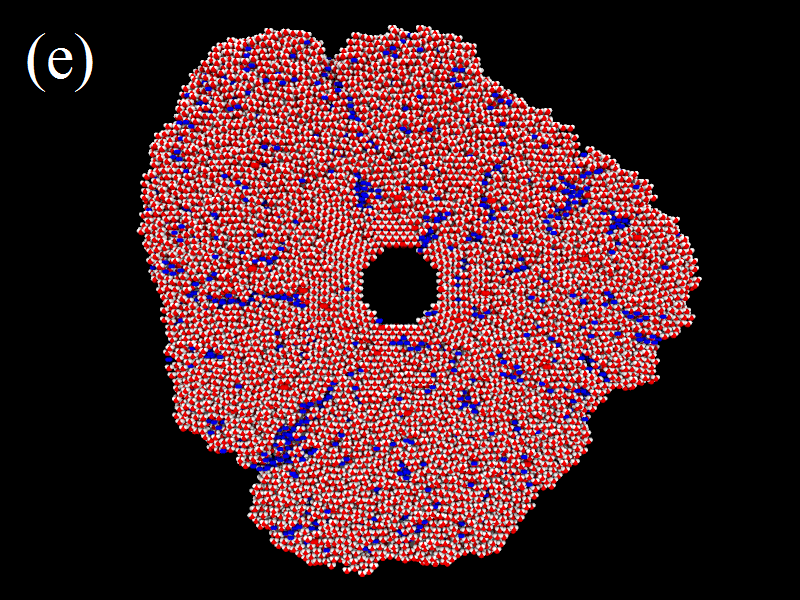}
\caption{\label{f_cross} Cross-sections of the simulated dust-grain ice mantles shown in Fig. \ref{f_surface}. Each cross section is  3 \AA \, deep, and passes through the center of the grain at arbitrary angle; molecular hydrogen (H$_2$) is highlighted in blue to aid the eye. {\bf (a)} Mantle formed through direct accretion of H$_2$O, with no surface chemistry and arbitrary gas density; {\bf (b)} mantle formed through accretion and surface chemistry of H and O atoms, with a gas density of $n_{\mathrm{H}} = 2 \times 10^{7}$ cm$^{-3}$; {\bf (c)} $n_{\mathrm{H}} = 2 \times 10^{6}$ cm$^{-3}$ ; {\bf (d)} $n_{\mathrm{H}} = 2 \times 10^{5}$ cm$^{-3}$ ; {\bf (e)} $n_{\mathrm{H}} = 2 \times 10^{4}$ cm$^{-3}$. {\em Videos of panels (a), (b) and (e) may be found in the online version of the journal, showing each cross section through 360 degrees.}}
\end{figure*}

\begin{figure*}
\center
\includegraphics[width=0.8\textwidth]{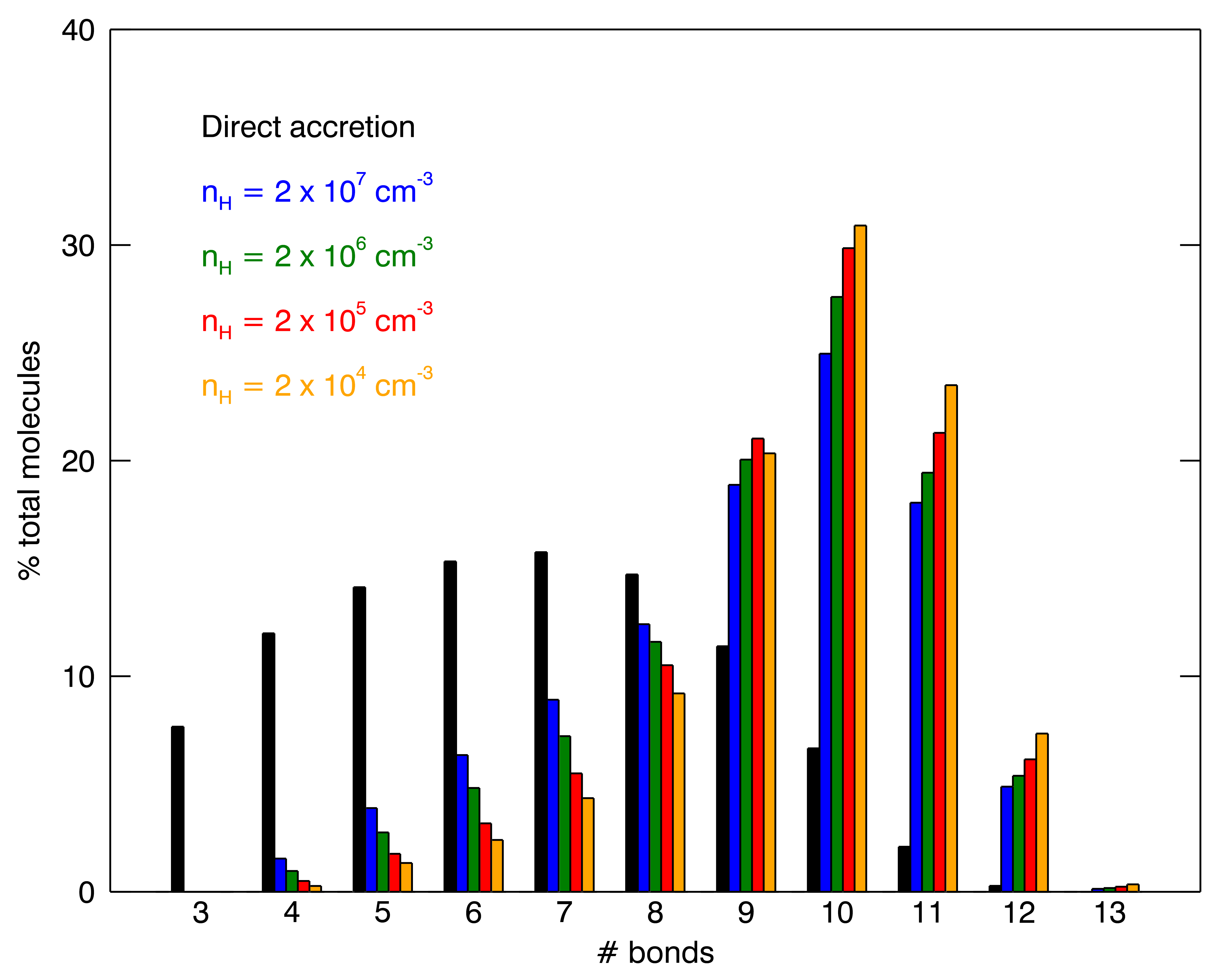}
\caption{\label{bar_coord} The percentage of all molecules in each model, binned by the number of binding partners, i.e. nearest neighbors. A single random-seed run is shown in each case. Direct accretion of water molecules (black bars) shows a qualititatively different binding distribution from models in which the ice mantle is formed by active surface chemistry (all other colors).}
\end{figure*}

\begin{figure*}
\center
\includegraphics[width=0.8\textwidth]{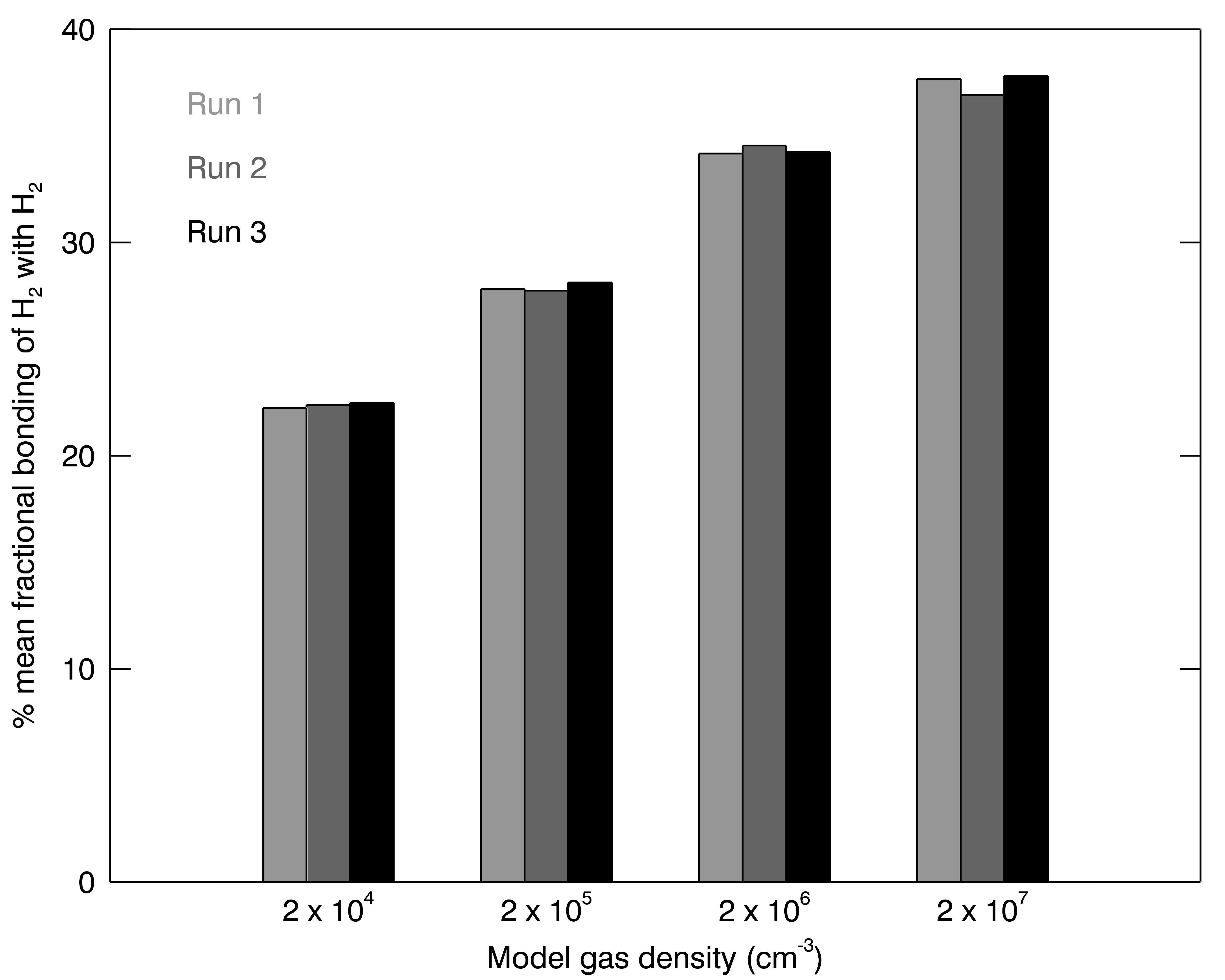}
\caption{\label{bar_H2} The number of H$_2$ molecules to which each H$_2$ molecule is bound, expressed as a percentage fraction of the number of bonds per molecule, averaged over all H$_2$ molecules. Higher-density models show somewhat stronger clustering of H$_2$ molecules, but the amount of H$_2$--H$_2$ bonding for all models is above the statistical expectation.}
\end{figure*}

\begin{figure*}
\center
\includegraphics[width=0.48\textwidth]{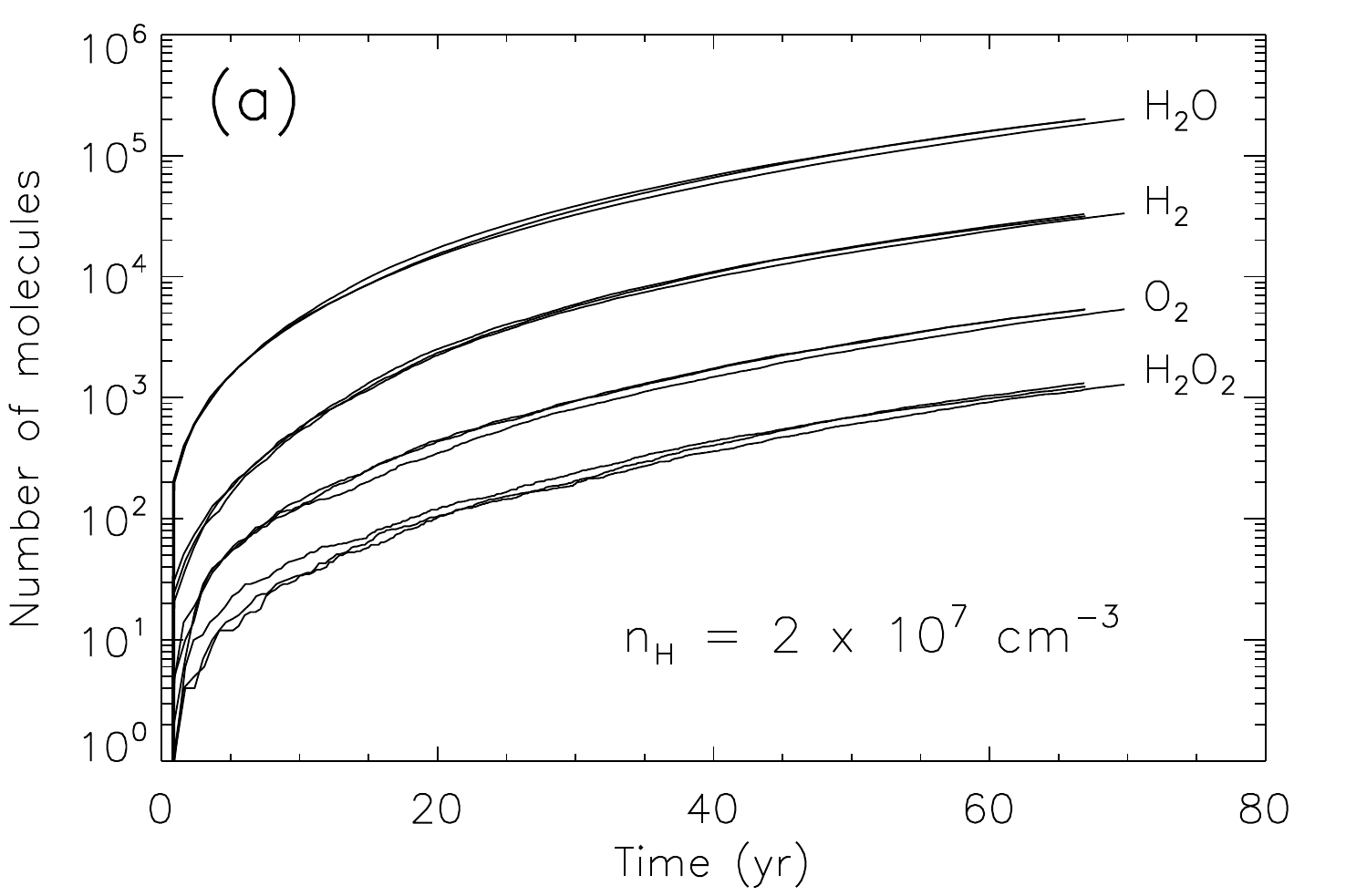}
\includegraphics[width=0.48\textwidth]{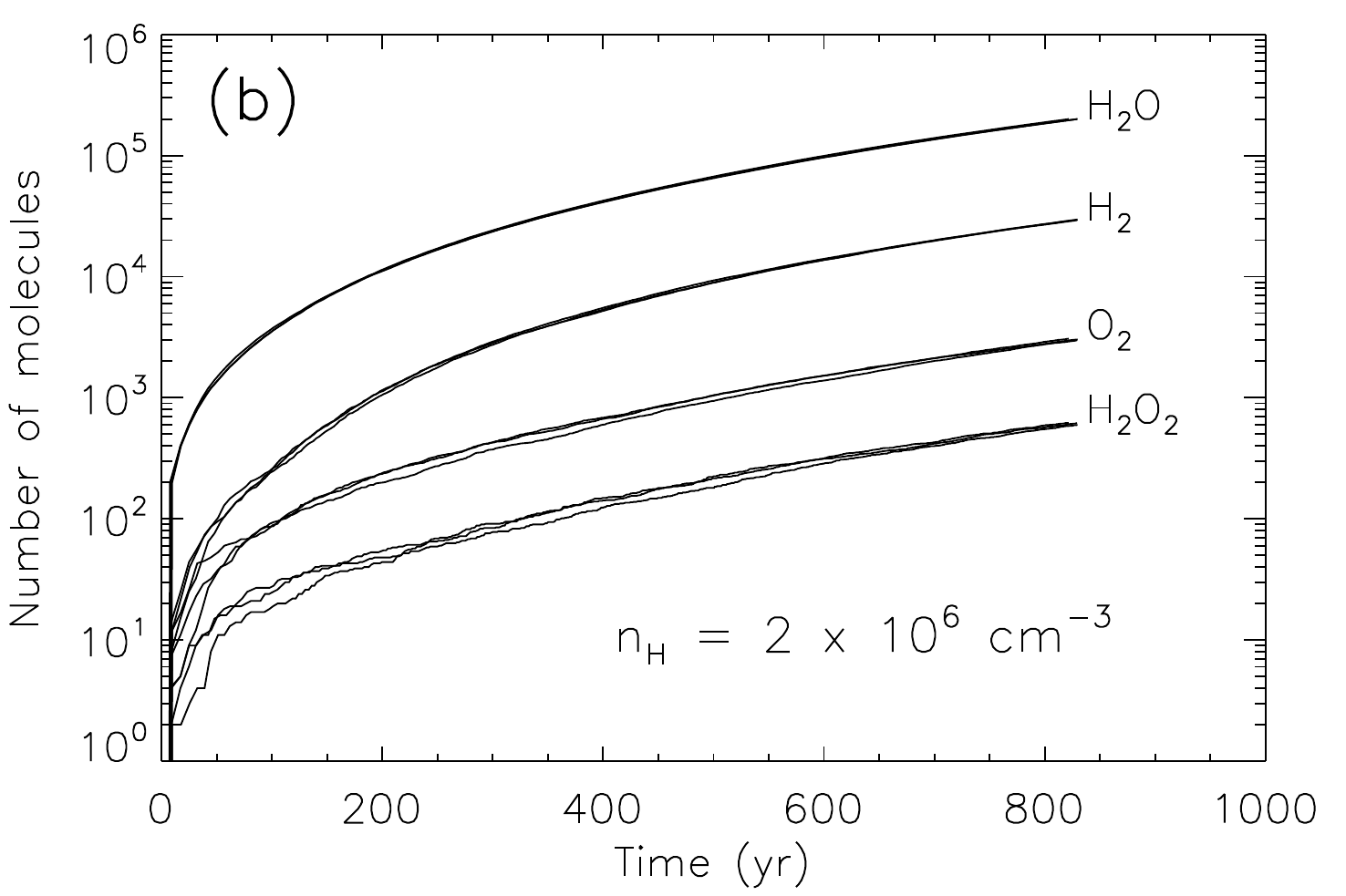}
\includegraphics[width=0.48\textwidth]{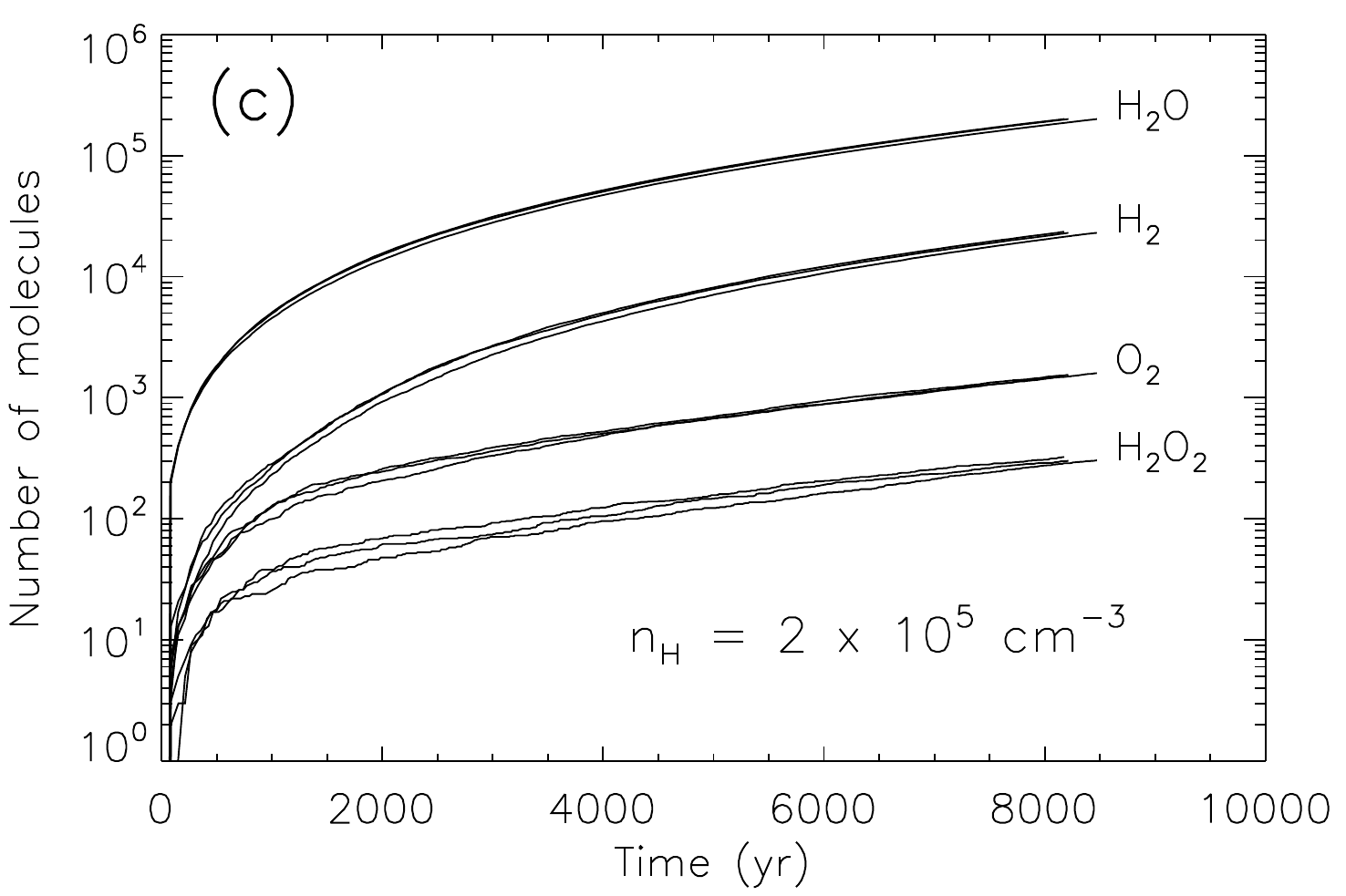}
\includegraphics[width=0.48\textwidth]{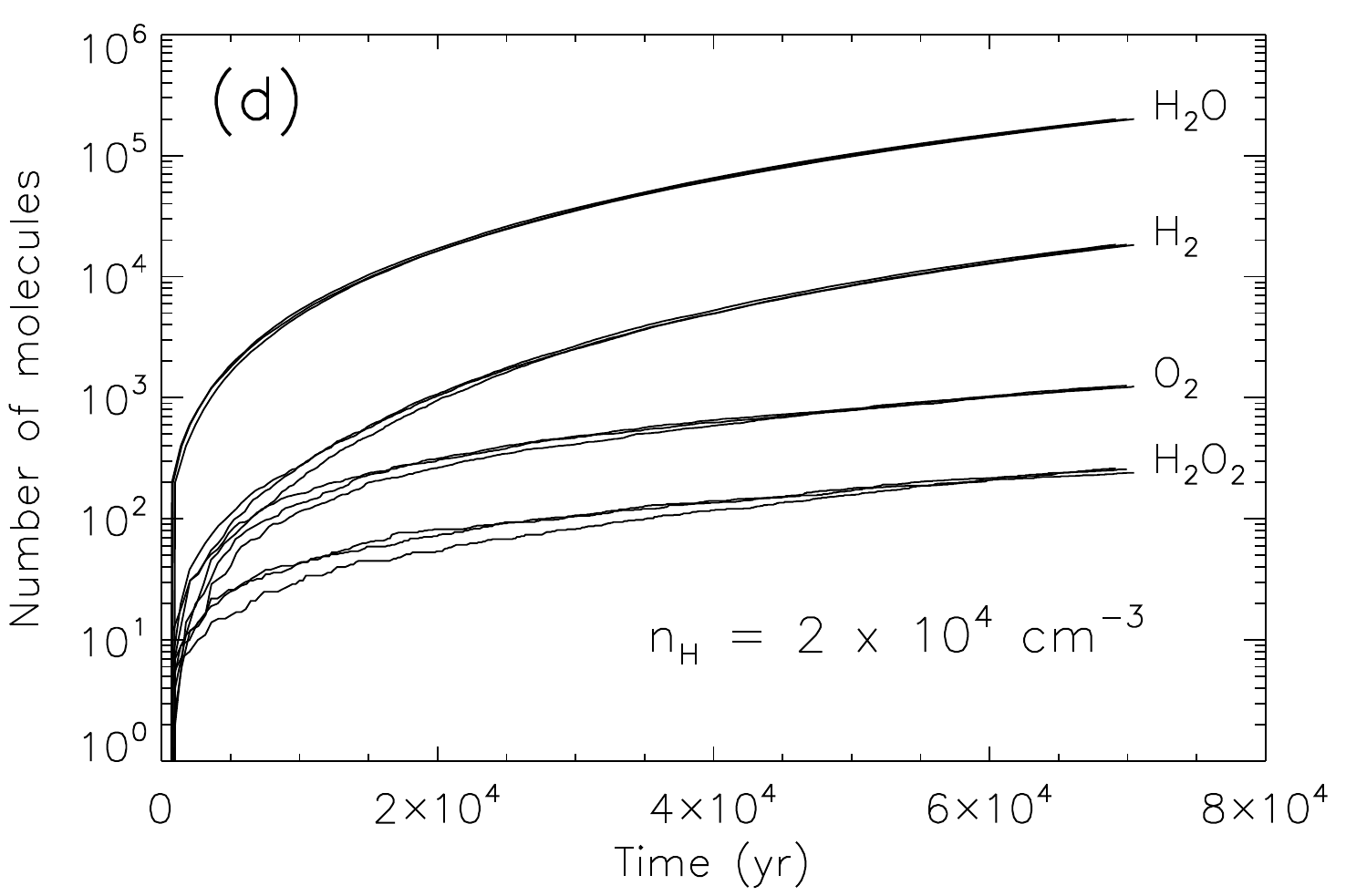}
\caption{\label{fig-abuns} Time-dependent abundances of stable species on the dust grain. Each panel corresponds to a different gas-density, showing results from three identical models using different random-number seeds.}
\end{figure*}


\end{document}